\documentclass[
reprint,
superscriptaddress,
 amsmath,amssymb,
 aps,
 prx
]{revtex4-2}

\usepackage{graphicx}
\usepackage{dcolumn}
\usepackage{bm}
\usepackage{xcolor}
\usepackage[normalem]{ulem}
\usepackage[english]{babel}
\usepackage{braket}
\usepackage{units}
\usepackage[T1]{fontenc}
\usepackage{titletoc}
\usepackage{physics}
\usepackage{comment}

\usepackage[colorlinks]{hyperref} 
\hypersetup{
	colorlinks=true,
	linkcolor=blue,
	filecolor=blue,      
	urlcolor=blue,
	citecolor=blue
}

\usepackage{ifthen}
\newboolean{showcomments}
\setboolean{showcomments}{true}

\makeatletter
\newcommand{\mynote}[3]{%
  \ifthenelse{\boolean{showcomments}}{%
   \fbox{\bfseries\sffamily\scriptsize#1}%
   {\small$\blacktriangleright$\textsf{\emph{\color{#3}{#2}}}$\blacktriangleleft$}}%
  {%
   \@bsphack
   \@esphack
  }%
}
\makeatother

\newcommand{\Eq}[1]{Eq.\,(\ref{#1})}

\newcommand{\Fig}[1]{Fig.\,\ref{#1}}

\newcommand{\gr}{\mathbf{G}}

\newcommand{\UI}{\bm{u}}
\newcommand{\DU}{D}

\newcommand{\FSen}{\bm{\theta_{ \rm en}}}
\newcommand{\FSde}{\bm{\theta_{\rm de}}}
\newcommand{\Uen}{\mathcal{U}_{\rm {en}}}
\newcommand{\Uem}{\mathcal{U}_{\rm {em}}}
\newcommand{\Ude}{\mathcal{U}_{\rm {de}}}
\newcommand{\EUI}[1]{\mathbb{E}_{\bm{u}}\!\!\left[#1\right]}

\newcommand{\ci}{\mathbf{V}}       

\begin{document}

\title{End-to-end Optimization of Single-Shot Quantum Machine Learning for Bayesian Inference}

\author{Theodoros Ilias}
\affiliation{Department of Electrical and Computer Engineering, Princeton University, Princeton, NJ 08544, USA}

\author{Fangjun Hu}
\affiliation{Department of Electrical and Computer Engineering, Princeton University, Princeton, NJ 08544, USA}

\author{Marti Vives}
\affiliation{Department of Electrical and Computer Engineering, Princeton University, Princeton, NJ 08544, USA}

\author{Hakan E. T\"ureci}
\affiliation{Department of Electrical and Computer Engineering, Princeton University, Princeton, NJ 08544, USA}

\date{\today}

\begin{abstract}
We introduce an end-to-end optimization strategy for quantum machine learning that directly targets performance under finite measurement resources, where learning objectives are defined directly at the level of task performance. The method is applied on a Bayesian quantum metrology task since it provides a natural testbed with known fundamental limits and scaling with system size. The sampling-aware hybrid algorithm achieves a single-shot risk within 1 dB of the -20 dB Bayesian limit using 32 qubits. We extend the Bayesian framework from parameter estimation to global function inference, where the task is to infer a target function of the sensor input drawn from an arbitrary prior, and we demonstrate a clear computational–sensing advantage for direct functional inference over indirect reconstruction. We relate the corresponding Bayesian risk to the Capacity metric and argue that the Resolvable Expressive Capacity provides a natural measure of the space of functions accessible in a single shot. The resulting eigentask analysis identifies noise-robust feature combinations that yield compact estimators with improved accuracy and reduced optimization cost in resource-limited or real-time on-device settings. Going beyond the irreducible statistical uncertainty associated with finite measurement records, we study implementation-induced distortions of the sensor response focusing on representative imperfections arising from readout errors and static coherent gate disorder. We show that when these imperfections are stationary across training and inference, end-to-end optimization can adapt the quantum circuit and classical estimator to the implemented hardware response, preserving high inference performance over a broad range of imperfection strengths.

\end{abstract}

\maketitle

\section{Introduction}

Training and inference on quantum hardware are fundamentally constrained by finite coherence times, limited measurement resources, restricted physical encodings, and deviations between the physical hardware and its associated idealized numerical model. In this work, we focus primarily on the limitations imposed by finite sampling, which constitute a dominant and inescapable bottleneck for near-term quantum machine learning (QML). A common approach in QML is to optimize a target loss under idealized, effectively infinite-shot assumptions, and only subsequently assess performance when a finite number of measurements is available. This separation has two important drawbacks. First, it obscures the measurement cost of implementation: features that are useful in the infinite-sampling limit may be poorly resolved at finite shot budgets, so that realizing the nominal performance can require a prohibitively large number of measurements~\cite{PhysRevX.13.041020}. Second, finite sampling modifies the effective learning problem itself, so that circuit parameters optimized in the infinite-shot limit need not remain optimal when the measurement budget is constrained. In addition, shot noise in finite-sample gradient estimates can, under suitable conditions, help gradient-based variational algorithms escape strict saddle points~\cite{PhysRevA.111.052441}. More broadly, these considerations align with calls to assess QML through operationally meaningful resources and task-level performance rather than computational separation alone~\cite{PRXQuantum.3.030101}. In the present work, finite sampling is built directly into the measurement model and inference objective at fixed shot budget. We optimize the quantum circuit, estimator, and training procedure end to end at fixed measurement budget, so that performance is defined under the same resource constraints encountered at inference. 

Our approach builds on recent developments in physical neural network (PNN) theory~\cite{PhysRevX.13.041020, Hu2024} and adapts them to the optimization of qubit-based devices under strict measurement constraints. In the PNN framework, the physical system is treated as the computational substrate, and its resolvable input–output functions are characterized directly from measured data. This naturally mitigates model–hardware mismatch because the estimator is optimized for the response of the actual device rather than an idealized surrogate. As we show below (Section~\ref{sec:results}), using the device itself as the sampler allows the optimization to adapt to stationary imperfections in the implemented response.

In QML applications, a central challenge in evaluating the advantage conferred by the use of a quantum substrate over a classical one lies in the absence of a systematic framework to determine the fundamental limit allowed by quantum mechanics. Equally important is a methodology for systematically comparing its performance to that of a classical substrate under equivalent resource constraints. We find that the application of a QML-based framework to Bayesian quantum metrology (BQM) offers a rigorous basis to address such challenge because of (1) the existence of a cost function for which the scaling behavior and performance with system size can be derived and computed, and (2) a comparative analysis can in principle be conducted with respect to the equivalent class of sensors operating in the classical limit. We present one approach here that applies this general strategy.

From the perspective of quantum metrology, a PNN-based approach provides a data-driven methodology that adapts naturally to the control and measurement resources of a given physical sensor. Limitations in accessible resource states, available Hamiltonian interactions, and readout fidelity often create bottlenecks that require sophisticated techniques to approach Heisenberg scaling~\cite{Giovannetti1330,PhysRevLett.96.010401, RevModPhys.90.035006}, the ultimate limit accessible through linear quantum measurements, as demonstrated in recent work utilizing collective-state protocols without single-particle detection \cite{PhysRevLett.116.053601}, squeezed superradiance-based readout schemes \cite{PhysRevLett.131.060802}, and scrambling-enhanced metrology~\cite{doi:10.1126/science.adg9500}. A PNN framework provides a complementary route by optimizing directly within the device’s physical constraints, using the data generated by the sensor as it samples the signal. The effectiveness of such an approach has been demonstrated with variational quantum metrology~\cite{PhysRevX.11.041045,PRXQuantum.4.020333}, that have the utility for both preparing metrologically-useful entangled resource states, and also in producing decoders that are able to effectively concentrate
the parameter-dependence of the entangled state into simple observables that can be easily measured. A recent experiment provides a proof-of-principle demonstration in a trapped-ion platform that this approach can be highly effective~\cite{Marciniak2022}. 

The task considered here is to train a parametrized circuit to estimate a target function, $f^*(u)$, of an input phase variable $u$ in a single measurement. The input is drawn from a prior distribution, $p(u)$, and the cost function maximizes the accuracy of the inferred function value. This formulation extends Bayesian quantum metrology~\cite{PhysRevX.11.041045,Marciniak2022,PRXQuantum.4.020333} to global function estimation under arbitrary priors, a task we refer to as {\it Bayesian quantum inference} (BQI).

A principled comparison across physical substrates requires a metric that quantifies the space of functions that can be resolved with fixed resources. The Resolvable Expressive Capacity (REC) introduced in Ref.~\cite{PhysRevX.13.041020} provides such a measure. For a single-shot budget it reduces, for linear function targets and Gaussian priors, to the Bayesian mean-square error used in single-parameter metrology~\cite{PhysRevX.11.041045,Marciniak2022} (see Section~\ref{sec:metrology_capacity}). REC therefore offers a natural extension of this metric to the broader BQI setting. For qubit-based systems under projective measurements, REC admits a tight analytic bound~\cite{PhysRevX.13.041020}. A spectral analysis of REC identifies eigentasks, the functions that can be approximated with the highest signal-to-noise ratio~\cite{PhysRevX.13.041020, hu_generalization_2024}. These provide the smallest latent space of effective features and can drastically reduce the optimization burden compared to working in the exponentially large full feature space.

The measurement-driven optimization of sensor parameters for single-shot estimation is expected to be a challenging problem. In the asymptotic limit of infinite measurement shots, where output features converge to their quantum expectation values and sampling noise vanishes, gradient-based variational QML approaches have been shown to scale poorly due to vanishing gradients~\cite{mcclean_barren_2018}, rendering gradient-based training practically ineffective due to the finite number of experiments that can be practically performed. Although circuit expressivity theoretically increases with depth~\cite{schuld_effect_2021, wu_expressivity_2021, du_efficient_2022, Harrow2009}, the optimization landscape of sufficiently deep parameterized quantum circuits exhibits exponentially vanishing gradients, leading to the well-known barren plateau phenomenon \cite{mcclean_barren_2018}. These results have motivated the adoption of gradient-free hybrid optimization methods, which, although they do not circumvent the associated concentration of the cost landscape \cite{Arrasmith_2021}, can nevertheless be advantageous in practice by avoiding explicit gradient estimation and by facilitating parameter-transfer or warm-start strategies across related problem instances. A class of QML algorithms well suited to NISQ devices that avoids the optimization challenges of variational approaches is quantum reservoir computing (QRC)~\cite{dambre_information_2012, fujii_harnessing_2017, chen_temporal_2020, wright_capacity_2019, garcia-beni_scalable_2022, Kalfus_2022}. When combined with mid-circuit feature extraction under finite-shot constraints, QRC can address limitations imposed by finite coherence times, as demonstrated by the NISQRC algorithm~\cite{Hu2024}. In QRC, an untrained quantum system serves as a high-dimensional nonlinear feature generator. Its fixed dynamics transform input data into a rich set of observables. A linear estimator in this space of non-linear features is optimized on labeled training data. Because only the classical weights are trained, a convex optimization efficiently solvable via standard linear algebra routines such as SVD, training requires far fewer quantum evaluations and avoids the vanishing-gradient issues that hinder variational approaches. Recent experimental demonstrations using atom-based reservoirs (108 qubits) \cite{kornjaca_large_2024} and Gaussian boson samplers (8176~\cite{gong_enhanced_2025} and 400~\cite{cimini_large_2025} modes, respectively), have highlighted the potential for scalability of this approach. However, these studies also report a rapid loss in accuracy when the sampling is reduced \cite{garcia-beni_scalable_2022, havlicek_supervised_2019, shen_deep_2017}. 

In the present work, we go beyond standard QRC by also optimizing the internal circuit parameters, while retaining a reservoir-inspired linear estimator. We show that training the internal parameters directly under finite-shot constraints substantially improves inference accuracy and reduces the required measurement budget to the single-shot regime for the BQI tasks considered. For the circuit family studied here, these results further suggest that high-performance BQI solutions remain accessible under direct on-device optimization.

To this end, we introduce a sampling-aware, end-to-end hybrid optimization algorithm for single-shot BQI in which state preparation, encoding, measurement, and classical post-processing are optimized jointly. This unified treatment contrasts with approaches that analyze performance through Fisher-information bounds \cite{Kay97,Helstrom1969,PhysRevLett.72.3439}, focus primarily on the design of probe states \cite{PhysRevLett.123.260505,Pedrozo-Penafiel2020,Colombo2022}, or optimize the final measurement \cite{Koczor_2020,Meyer2021,PhysRevLett.128.160505,MacLellan2024,PhysRevA.111.042432}. These perspectives often address individual elements of the sensing pipeline, whereas our framework directly targets the overall inference task and the estimator that ultimately determines performance in the single-shot regime. Applied to a single-parameter metrology task, the proposed method reaches a risk of -19.1 dB with 32 qubits, close to the -20 dB limit achieved by the optimal Bayesian sensor under the same resources~\cite{Macieszczak_2014}. Unlike that optimal sensor, our circuit depth is fixed throughout the optimization.

Recent work has introduced the concept of a quantum computational–sensing advantage~\cite{allen2025quantumcomputingenhancedsensing,khan2025quantumcomputationalsensingadvantage}, referring to potential performance gains achieved when quantum sensors are integrated with on-device quantum information processing to perform part of the inference task directly.

Importantly, this notion of advantage does not rely on reaching system scales where classical computation of equivalent accuracy becomes intractable. Here, we propose quantitative way to assess such comparative advantage based on the Bayesian risk under identical resource constraints: (i) a quantum sensor optimized to estimate a latent parameter, followed by classical post-processing to compute a desired function of that parameter; and (ii) a quantum sensor optimized end-to-end to infer that function directly in a single shot. The corresponding Bayesian risks for phase-based BQI in a 32-qubit system are found to be $-2.9$ dB in scenario (i) versus $-4.7$ dB in scenario (ii), respectively -- a statistically significant improvement that illustrates how direct functional optimization, i.e. BQI, can realize a measurable computational–sensing benefit (Section~\ref{sec:results}).

\begin{figure}[t]
    \centering
    \includegraphics[scale=1.0]
    {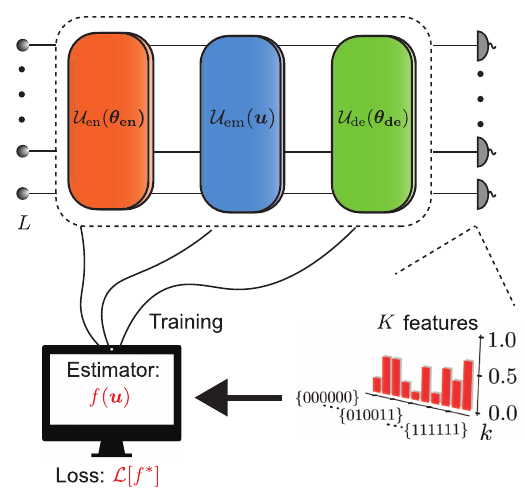}
    \caption{Physical neural network approach to quantum sensing. $L$ initially uncorrelated qubits are prepared in a metrologically useful state $\rho_{\rm en} = \Uen(\FSen)\rho_0$ via a parameterized entangling superoperator $\Uen(\FSen)$. In turn, the signal input $\UI$, drawn from a training dataset, is embedded on the quantum system. A parameterized decoder $\Ude(\FSde)$ is applied to the encoded state, followed by projective measurements in the computational basis. Collecting the measurement features allows to construct an estimation $f(\UI)$ of the target function $f^*(\UI)$ and calculate the objective loss function. Optimizing such loss function allow us to train the internal parameter of the quantum circuit as well as the estimator enabling the optimization of the whole sensing pipeline. We note that although our framework places no restriction on the form of the entangling, embedding and decoding operations and allows them to be arbitrary quantum maps, in present work we focus on unitary implementations.  } 
    \label{fig:Figure1}
\end{figure}

Beyond the irreducible statistical uncertainty of finite measurement records, we also examine implementation-induced distortions of the sensor response. We consider two physically distinct cases: readout errors, which alter the observed outcome distribution, and static coherent gate disorder, which modifies the underlying quantum dynamics. When these imperfections are stationary between training and inference, end-to-end optimization can adapt both the circuit parameters and the classical estimator to the modified hardware response, maintaining high inference performance over a broad range of imperfection strengths (see Section~\ref{sec:robustness_noise}). This should be interpreted as adaptation to a stationary, calibratable hardware map. In particular, we show that static coherent imperfections need not be intrinsically detrimental to reservoir-based inference.

The rest of the paper is organized as follows. In Section~\ref{sec:previous_work} we review the conceptual foundations of BQM while Section~\ref{sec:PNN_metrology} presents the general framework of our quantum computation  sensing approach and establishes its connection to the Resolvable Expressive Capacity (REC) introduced in Ref.~\cite{PhysRevX.11.041045} for finite-sample scenarios. In Section~\ref{sec:resevoir_optimization}, we detail an efficient optimization procedure that is essential for the successful end-to-end training of the sensing pipeline. Section~\ref{sec:use_of_eigentasks} introduces the concept of eigentask learning with the robustness to various imperfections being studied in Section~\ref{sec:robustness_noise}, while Section~\ref{sec:arbitrary_tasks} provides several illustrative examples of approximating functions of inputs drawn from various probability distributions, thereby highlighting the versatility and advantages of our approach. Finally, Section~\ref{sec:conclusions} summarizes our findings and outlines potential directions for future research.

\section{Theory \& Results}
\label{sec:theory_results}

\subsection{Review of Bayesian Quantum Metrology}
\label{sec:previous_work}

Here, we review the Bayesian quantum metrology framework for single-parameter estimation~\cite{PhysRevX.11.041045,Marciniak2022}, including the task-specific loss function, known results on optimal sensor configurations, and the corresponding quantum limits. This provides the basis for our extension to global function estimation under arbitrary priors. For concreteness, we focus on Ramsey interferometry, where performance is quantified by single-shot SNR for phase estimation.

Consider the estimation of a phase $u$ imparted by a weak electromagnetic pulse on a qubit-based sensor system composed of $L$ qubits. The pulse whose properties one wants to estimate encodes a phase in the sensor by evolving to the state $\rho_{u} = \Uem(u)\rho_{\rm en}$, where $\rho_{\rm en}$ is a prepared probe state and $\Uem(u) \cdot \equiv e^{-iu\hat{H}} \cdot e^{iu\hat{H}} $ the unitary embedding channel generated by $\hat{H}$ and encodes the unknown phase $u$. Estimation of $u$ is performed using measurement outcomes $j$, defined by a positive operator-valued measure (POVM) $\{ \hat{M}_j \}_{j=1}^K$. Conditioned on the true value of $u$, outcome $j$ occurs with probability $p(j|u) = x_j(u) = \operatorname{tr}\left\{ \hat{M}_j \rho_u \right\}$. A natural measure of estimation accuracy is the mean-squared error (MSE) ${\rm MSE}(u) =\sum_{j=1}^K (u-u_{\rm est}(j))^2 p(j|u)$ with $K$ being the number of the possible outcomes. Minimizing such a metric simultaneously for a wide range of phases $u$ is impossible since probe states, POVMs and estimators that are optimal for some phase values might not work for some others~\cite{paris_2009}. Consequently, a more appropriate metric corresponds to the MSE averaged over the prior distribution of $u$, $p(u)$ that reflects the statistical properties of the unknown parameter suggests the use of BMSE, ${\rm BMSE}=\int du\, p(u) \, {\rm MSE}(u)$~\cite{PhysRevX.11.041045}. 

Although lower bounds on the BMSE described above optimized over all possible probe states, estimators and POVMs together with the construction of an iterative numerical algorithm to find them has been studied in the past \cite{Macieszczak_2014,Jarzyna_2015,PhysRevLett.124.030501}, it has only recently been shown that one can approach such bounds using shallow circuits composed of experimentally accessible gates, both in the single-parameter \cite{PhysRevX.11.041045,Marciniak2022} and multi-parameter \cite{PRXQuantum.4.020333} settings. 

\subsection{PNN approach to Global Function Sensing}
\label{sec:PNN_metrology}

In this section, we define the six components of the quantum computational-sensing framework that together specify the end-to-end BQI architecture used in the metrological setting considered below. The formulation below addresses the general case of an $S$-shot budget and high-dimensional inputs $\boldsymbol{u} = (u_1, \ldots, u_d)$.

\begin{enumerate}
    \item \textbf{Training/Testing Dataset:} We assume access to a labeled dataset of size $N_{\rm tot}$ consisting of input–output pairs $\{ \UI_i, \bm{f}^*(\UI_i) \}$, where $\bm{f}^* = (f_1, \ldots, f_{D})$ is the vector-valued target function to be learned and the inputs $\UI$ are sampled from a general prior distribution $p(\UI)$ with a compact support. This formulation naturally accommodates multi-parameter estimation. Following standard machine-learning practice, we randomly partition the dataset into a training set of size $N_{\rm tr}$, used for optimizing the model parameters, and a testing set of size $N_{\rm ts}=N_{\rm tot}-N_{\rm tr}$, used exclusively to evaluate the generalization performance of the trained model.

    \item \textbf{Entangled Probe Preparation:} The system is initialized in a fixed quantum state $\rho_0$ and subsequently transformed into a metrologically useful state $\rho_{\rm en} = \Uen(\FSen)\rho_0$ via a parameterized entangling superoperator $\Uen(\FSen)$, where $\FSen$ denotes a set of tunable parameters. Henceforth, the calligraphic operators refer to superoperators defined as $\mathcal{U}\cdot=U \cdot U^{\dagger}$.

    \item \textbf{Input Embedding:} The classical input $\UI$ is embedded into the quantum state using a parameter-dependent superoperator $\mathcal{U}_{\rm em}(\UI)$, yielding the input-encoded state $\rho_{\rm em}(\UI) = \mathcal{U}_{\rm em}(\UI)\rho_{\rm en}$.

    \item \textbf{Decoding and Feature Extraction via Quantum Measurement:} A parameterized decoder $\Ude(\FSde)$ is applied to the encoded state, followed by projective measurements in the computational basis. This measurement process realizes a positive operator-valued measure (POVM) $\{ \hat{M}_j \}_{j=1}^K$, allowing access to information encoded in the final state
    $$\rho(\UI) = \Ude(\FSde) \, \mathcal{U}_{\rm em}(\UI) \, \Uen(\FSen)\rho_0.$$
    
    Performing $S$ repeated measurements (shots) on this state yields a sequence of independent outcomes ${\bm{j}}= (j_1, j_2, \ldots, j_S)$, from which we construct the empirical \textit{feature vector} $\mathbf{X}(\UI) = (X_1(\UI), X_2(\UI), \ldots, X_K(\UI))$ with $X_j = \frac{1}{S} \sum_s \delta(j_s - j)$. Thus $X_j$ is the empirical $S$-shot frequency of outcome $j$ and can be extracted from a physical device by performing $S$ experiments for any input $\UI$ sampled from $p(\UI)$ and vectorizing the histogram of outcomes (see \Fig{fig:Figure1}). Equivalently, the parameters $X_j$ correspond to the maximum-likelihood estimates of the categorical distribution over $K$ outcomes, where the conditional probability of obtaining outcome $j$ given the control $\UI$ is $p(j \mid \UI) = x_j(\UI) = \operatorname{tr}\!\left\{ \hat{M}_j\, \rho(\UI) \right\}$, as inferred from an $S$-shot experiment. In what follows, it will be useful to define the $S$-shot fluctuation vector by subtracting the expectation $X_j(\UI) = x_j(\UI) + \zeta_j(\UI)$, where now $\zeta_j(\UI)$ denotes a zero-mean stochastic fluctuation due to finite-sample (quantum shot) noise characterized with covariance $\mathrm{Cov} [\bm{\zeta} (\bm{u})] = \frac{1}{S} \bm{\Sigma} (\bm{u})$, where $\bm{\Sigma} (\bm{u}) = \mathrm{diag} (\bm{x} (\bm{u})) - \bm{x} (\bm{u}) \bm{x} (\bm{u})^T$. For additional details, see Ref.~\cite{PhysRevX.13.041020}.

    \item \textbf{Estimator Construction and Training:} Given the measured features $\mathbf{X}(\UI)$, we approximate the target function using a linear estimator of the form
    \begin{align}
    f_{\mathrm{est,i}}(\UI) = \sum_{j=1}^K w_{ij} X_j(\UI),
    \label{eq:estimator}
    \end{align}
    where $w_{ij}$ is a learnable weight matrix with dimension $D \times K$. This estimator based on a complete set of POVMs was introduced in Ref.~\cite{PhysRevX.13.041020} for characterizing the function expression capacity of a given physical system under finite sampling of its projectively measured outputs. Here, we adopt this estimator because (1) it cleanly separates the non-linear action of the sensor embodied in its measured features $X_k(\UI)$ from a postprocessor that has a minimal action -- a linear projection, and (2) because it allows the hardware-efficient training of the sensor parameters $\left( \FSen, \FSde \right)$ through an optimization methodology we introduce in Section~\ref{sec:resevoir_optimization}.

\item \textbf{The Cost Function:} A key quantity of our analysis is the definition of the cost function which in general takes the form:
\begin{align}
    l[f^*] =  \EUI{ \mathbb{E} \left[ \left( \bm{f}^*(\UI) - \bm{f}_{\mathrm{est}}(\UI) \right)\cdot \left( \bm{f}^*(\UI) - \bm{f}_{\mathrm{est}}(\UI) \right) \big| \UI \right]}.
    \label{eq:loss}
\end{align}
Here, $\EUI{g(\UI)} \equiv \int d\UI \, p(\UI) \, g(\UI)$ denotes the expectation over the input distribution $p(\UI)$, and $\mathbb{E}[\,\cdot\,| \UI]$ represents the expectation over measurement outcomes. Note here the similarity of such quantity to the BMSE for the special case of $f^*(u)=u$. For $S$-shot sampling of POVM measurements with outcomes $\bm{j}= (j_1, j_2, \ldots, j_S)$, it is formally defined by  $\mathbb{E}[\,\cdot\,|\UI]\equiv \sum_{\bm{j}} p({\bm{j}}|\UI)\, [\,\cdot\,]$. Here, the sum is taken over all possible $S$-measurement outcome vectors $\bm{j}$. If the measurements are independent of each other, $p(\bm{j}|\UI)$ only depends on the count vector $\bm{N} (\bm{j}) = (N_1,\dots,N_K)$ whose components are non-negative integers satisfying $\sum_{j=1}^K N_j = S$, and $P(\bm{N}(\bm{j})|\UI) = \frac{S!}{\prod_{j=1}^K N_j!} \prod_{j=1}^K p(j|\UI)^{N_j}$, where $p(j|\UI)$ is, as defined earlier through the Born rule, the probability of obtaining outcome $j$ in an experimental run for the given input $\UI$. Furthermore, adopting an estimator such as in Eq.~\eqref{eq:estimator} and minimizing the loss function with respect to the weight vector, $\bm{w}$, we arrive to the definition of the post reservoir loss function
\begin{equation}
    {\mathcal{L}}(\bm{\theta}) = \min_{\bm{w} \in \mathbb{R}^K} \, {l}(\bm{\theta}, \bm{w}), 
    \label{eq:post_reservoir_loss_inf}
\end{equation}
where $\bm{\theta}$ denotes all the internal parameters. The role of this loss function is discussed further in Sec.~\ref{sec:resevoir_optimization}. The framework applies generally to mappings $f:\mathbb{R}^d\rightarrow\mathbb{R}^D$, accommodating multi-parameter inputs and vector-valued targets. For example, estimating $B_x^2+B_y^2+B_z^2$ from the field components $(B_x,B_y,B_z)$ corresponds to $d=3$ and $D=1$. In the remainder of this work, we restrict attention to the scalar case $d=D=1$, i.e., $f:\mathbb{R}\rightarrow\mathbb{R}$.
Finally, in appendix \ref{app:Fisher_Bhattacharyya} we show that in the special case of a single shot estimation ($S=1$) of a scalar parameter $u$ drawn from a Gaussian prior $p(u)$, the BQI cost function coincides with the Bayesian MSE~\cite{vasilyev_optimal_2024}
when the target function is chosen to be a linear function $f^* (u) = u$:
\begin{equation}
     \mathcal{L} [f^*] \rightarrow \int du \, p(u) \sum_{j=1}^{K} p(j|u) \, (u - u_\text{est} (j))^2 
\end{equation}
where the estimator is given by $u_\text{est} (j) = \sum_\ell w_{\ell} \delta_{\ell,j}$.

\end{enumerate}

\subsubsection{Single-parameter Metrology and Capacity}
\label{sec:metrology_capacity}

The cost function defined in Eq.~\eqref{eq:post_reservoir_loss_inf} serves as the central performance metric in our data-driven sensing framework, quantifying how well the optimized sensing pipeline approximates a target function $f^*(u)$. Beyond the role as an optimization objective, we discuss in this section how this cost function can serve as a metric for characterizing the space of functions that a given sensor can resolve with the optimal estimator \Eq{eq:estimator}. The cost function Eq.~\eqref{eq:loss} is linearly related to the \textit{Capacity} $C[f^*]$ introduced in Ref.~\cite{PhysRevX.13.041020} 
\begin{align}
    C[f^*] = 1 - \frac{\mathcal{L} [f^*]}{\mathbb{E}_{\UI}[f^{*2}]}
    \label{eq:fcap}
\end{align}
Normalization over the prior here ensures that $0 \le C \le 1$. This connection brings out three important observations. First, given a sensor with internal parameters $(\FSen, \FSde)$, a tight bound can be found for the total function expression capacity the sensor can resolve at $S$ shots through the estimator \Eq{eq:estimator}. This is a real number $C_T$ that can be estimated by experimentally sampling the physical system constituting the sensor in its input space subject to the prior distribution $p(u)$. Second, a set of $2^L$ orthonormal functions $y^{(j)}$ can be constructed that have provably highest SNR at $S$ shots. These results will be discussed in Sec.~\ref{sec:eigentasks}. Third, as we show in this section, when the prior is broad over the support of $u$, the capacity $C$ serves as a global information-theoretic measure that, in the local limit, reduces to previously introduced information-theoretic quantities in the case of single-parameter metrology.  

The single-parameter sensing paradigm can then be equivalently recast as a function approximation task, where the goal is to learn the target function:
\begin{align}
    f^*(u)=u.
    \label{eq:target_sensing}
\end{align}
It is important to note that the sensing loss function, $\mathcal{L}[f^*]$ in Eq.~\eqref{eq:loss}, depends on the number of shots $S$ carried out to construct an estimate of the function. It notably depends on the prior probability distribution of the inputs $p(\UI)$. Therefore, let us choose the special case of single-shot estimation, i.e. $S=1$, and suppose that the input distribution $u$ obeys a zero-mean Gaussian distribution
\begin{align}
    p(u)=\frac{1}{\sqrt{2 \pi} \sigma} e^{-\frac{u^2}{2\sigma^2}}.
    \label{eq:Gaussian_distr}
\end{align}

Under these assumptions, the single-shot sensing loss associated with a given POVM ${\hat{M}_j}$ admits the expansion (see Ref.\,\cite{hu_thesis}, see also Appendix~\ref{app:Fisher_Bhattacharyya}):
\begin{align}
    \mathcal{L}[f^*;S=1] = \sigma^2 - \sigma^4 \mathcal{I}^{(1)}(0) - \sigma^6 \mathcal{I}^{(2)}(0) + \mathcal{O}(\sigma^8).
    \label{eq:linear_capacity}
\end{align}
The second-order term corresponds to the Fisher information (FI)~\cite{Kay97}, and can be expressed as
\begin{align}
    \mathcal{I}^{(1)}(0) = \sum_{j=1}^K \frac{\left( x_j^{(1)}(0) \right)^2}{x_j(0)},
    \label{eq:FI}
\end{align}
where $x_j^{(n)}(0)$ denotes the $n$-th order derivative with respect to the unknown parameter $u$ of the $j$-th measurement outcome evaluated at $u=0$. The third-order term in the $\sigma^2$ expansion is associated with what is known as \textit{Bhattacharyya information} \cite{a4026bed-4fc1-3dff-b24a-7f8d784a7c56}
\begin{align}
    \mathcal{I}^{(2)}(0) = \sum_{j=1}^K \left( \frac{x_j^{(1)}(0) x_j^{(3)}(0)}{x_j(0)}-\frac{(x_j^{(1)}(0))^2 x_j^{(2)}(0)}{2 (x_j(0))^2} \right).
    \label{eq:FI}
\end{align}
Consequently, the single-shot linear capacity can be viewed as a generating functional for a hierarchy of information measures: the Fisher information appears as the first-order contribution, followed by a series of higher-order Bhattacharyya-type terms.

These findings align with recent studies~\cite{e20090628,vasilyev_optimal_2024} showing that the Fisher information provides a reliable precision bound only in the local regime, where the parameter is confined to a narrow neighborhood around a known reference value. When the parameter is distributed over a broad range, the FI alone becomes insufficient to characterize the fundamental limits of estimation. In such \textit{global} scenarios, the higher-order Bhattacharyya information terms become essential, offering a more refined and accurate quantification of the system’s estimation capabilities as they account for both the nonlinear response of the measurement outcomes and the intrinsically nonlocal character of estimation when the parameter varies over a wide range. Optimizing over all possible POVMs elevates these classical quantities to their quantum counterparts, giving rise to the quantum Fisher information and the quantum Bhattacharyya or Barankin-type bounds~\cite{PhysRevLett.130.260801}. 
Consequently, the Capacity in Eq.~\eqref{eq:fcap} encapsulates all relevant information-theoretic contributions, both local and nonlocal, making it a natural figure of merit for sensing across all regimes. Its optimization not only recovers known bounds in the local limit but also may guide the design of improved measurement strategies in global scenarios and BQI-style tasks, where higher-order effects become essential.

\subsection{Reservoir-based optimization}
\label{sec:resevoir_optimization}

Optimizing the full sensing stack comprising the preparation of a metrologically useful state, the choice of POVM, and the classical post-processing used to compute the estimator presents substantial computational challenges. The parameter space is high dimensional, and the objective function $l[f^*]$ exhibits stochastic fluctuations under realistic conditions, particularly when the number of measurement shots $S$ is small relative to the number of possible outcomes $K$. In this regime, gradient-based optimization can become statistically inefficient and require a large number of shots.

In this regime, a carefully designed optimization strategy is essential, since joint training of the sensing components otherwise becomes trapped in suboptimal regions of the landscape. To overcome this difficulty, we develop a tailored optimization scheme suited to the complex, nonconvex structure characteristic of supervised machine learning problems, including our sensing task.

We consider a supervised quantum machine-learning setting in which classical inputs $\bm{u}$ are drawn from an unknown distribution $p(\bm{u})$. In practice, one has access to a finite training set consisting of $N_{\rm tr}$ independent and identically distributed (i.i.d.) input samples, $\{ \bm{u}^{(n)} \}_{n=1}^{N_{\rm tr}}$, drawn from $p(\bm{u})$, each paired with a corresponding target $f^*(\bm{u}^{(n)})$. The target may be continuous, as in regression tasks, or discrete-valued, as in classification problems.

As discussed previously, the post-processing stage involves a linear combination of features, written as $\bm{w}^T \bm{X}(\bm{u})$, where $\bm{w} \in \mathbb{R}^{K \times \DU}$ denotes the matrix of output weights and $\DU$ is the dimensionality of the output space.Therefore, we define the average loss over a dataset of $N$ samples as
\begin{align}
    \hat{l}(\bm{\theta}, \bm{w}) = \frac{1}{N} \sum_{n=1}^N \left( f^*(\UI^{(n)}) - f_{\mathrm{est}}(\UI^{(n)})) \right)^2,
    \label{eq:empir_loss}
\end{align}
as an estimator of Eq.\,(\eqref{eq:loss}). 
We note that in supervised learning, the loss function $\mathcal{L}$ also depends on the target label $f^*(\bm{u}^{(n)})$ associated with each input $\bm{u}^{(n)}$, although for notational simplicity this dependence is not written explicitly in the arguments of $\mathcal{L}$ throughout this article. 

When the internal parameters $\bm{\theta}$ are fixed after random initialization, training is restricted to the output weights $\bm{w}$. This corresponds to the standard \textit{reservoir computing} (RC) or \textit{extreme learning machine} paradigm, which has been extensively studied in both classical and quantum settings~\cite{GuangBinHuang, ortin_unified_2015, mujal_opportunities_2021, wright_capacity_2019, wilson_quantum_2019, innocenti_potential_2022}. A common choice for the reservoir readout is a linear estimator, $\bm{f}_{\rm est}=\bm{w}^{T}\bm{z}$, with the output weights trained by minimizing the mean-squared error, as in Eq.~\eqref{eq:loss}. For fixed features $\bm{z}$, the resulting empirical loss is a convex quadratic function of $\bm{w}$. The optimal weights $\bm{w}^{*}(\bm{\theta})$ can therefore be obtained efficiently by linear least-squares methods; the solution is unique when the corresponding feature Gram matrix has full rank, and otherwise can be defined through the minimum-norm pseudoinverse solution.

If one wishes to optimize both the internal and the external parameters a straightforward approach would be to optimize both $\bm{\theta}$ and $\bm{w}$ jointly by computing the gradients $\nabla_{\bm{\theta}} \hat{l}(\bm{\theta}, \bm{w})$ and $\nabla_{\bm{w}} \hat{l}(\bm{\theta}, \bm{w})$ and applying standard gradient descent. We show that it is advantageous to exploit the special structure of the problem, namely that the loss function $\hat{l}$ is convex with respect to $\bm{w}$.

To this end, we introduce a method that carries out gradient-based optimization directly on the post-reservoir loss
\begin{equation}
    \hat{\mathcal{L}}(\bm{\theta}) = \min_{\bm{w} \in \mathbb{R}^K} \, \hat{l}(\bm{\theta}, \bm{w}), 
    \label{eq:post_reservoir_loss}
\end{equation}
with the corresponding optimal weights $\bm{w}^*(\bm{\theta}) \equiv \arg\min_{\bm{w}} \, \hat{l}(\bm{\theta}, \bm{w})$. We can show that under reasonable assumptions, $\hat{\mathcal{L}}(\bm{\theta})$ can be approximated as a Gaussian random variable, $\hat{\mathcal{L}}(\bm{\theta}) \sim \mathcal{N}(\mu(\bm{\theta}), s(\bm{\theta}))$, see the detailed expression of $\mu(\bm{\theta}), s(\bm{\theta})$ in Appendix~\ref{appsec:cumulant-qsn}.

We seek an optimization strategy that leverages both the bounded search domain and the regularity properties of the stochastic loss function $\hat{\mathcal{L}}(\bm{\theta})$. For this purpose, we adopt a tailored variant of the \textit{Dividing RECTangles (DIRECT)} algorithm~\cite{Jones1993,article_direkt,nicholas2015dividing,article_direkt2}. This method admits global convergence guarantees under adequate refinement and is further adapted to incorporate a physically motivated statistical metamodel of the loss landscape, thereby enhancing convergence efficiency and mitigating sensitivity to local minima.

\begin{figure}[t]
    \centering
    \includegraphics[scale=1.0]
    {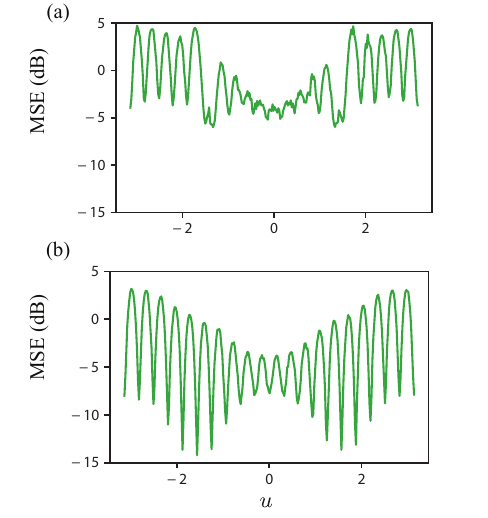}
    \caption{
Results of the optimization for approximating the target function $f(u)=\sin(10u)$ where $u$ is sampled from a Gaussian prior of width $\sigma=0.7981$ and for a circuit with $n_{\rm en}=1$, $n_{\rm de}=0$, $L=32$, and $S=1$. 
(a) Mean squared error, ${\rm MSE}(u)$, as a function of the true input phase $u$ by indirect approximation of the target function, achieving $\EUI{{\rm MSE_1}} \approx -2.9~{\rm dB}$
(b) Mean squared error, ${\rm MSE}(u)$, as a function of the true input phase $u$ by direct approximation of the target function achieving $\EUI{{\rm MSE_2}} \approx -4.7~{\rm dB}$ (see Sec.~\ref{sec:results} of the main text).
} 
    \label{fig:MSE_sin(10u)}
\end{figure}    

\subsection{Eigentasks analysis}
\label{sec:eigentasks}

Until now, our estimator has taken the form $f(\bm{u}) = \bm{w}^T \bm{X}(\bm{u})$, where $\bm{X}(\bm{u})$ denotes the empirical feature vector obtained from $S$ measurement shots. This representation implicitly includes a linear combination of all possible measurement outcomes. However, in practice, certain features—or specific linear combinations of features—may contribute disproportionately to the overall noise relative to the information they provide. Retaining such noisy components can degrade performance and might lead to overfitting.

To mitigate this, one can identify and eliminate redundant or noise-dominated features, akin to the spirit of principal component analysis (PCA), which selects only the most informative directions in feature space. In our context, these optimal linear combinations of features are referred to as \textit{eigentasks}, a concept first introduced in Ref.~\cite{PhysRevX.13.041020}, and shown to minimize generalization error in supervised machine learning tasks with large-sized training datasets \cite{hu_generalization_2024}.

The key equation underlying the definition of eigentasks is a generalized eigenvalue problem
\begin{align}
   \ci \bm{r}^{(j)} = \beta_j^2 \gr \bm{r}^{(j)}, 
   \label{eq:eigenprob} 
\end{align}
where $\gr \equiv \EUI{\bm{x} \bm{x}^T}$ and $\ci \equiv \EUI{\bm{\Sigma}}$ is the expected Gram and covariance matrix respectively. Each eigenvalue $\beta_j^2$ quantifies the MSE in constructing a system-specific function $y^{(j)} (\UI)$ given by 
\begin{align}
    y^{(j)} (\UI) = \sum_{j'} r_{j'}^{(j)} x_{j'} (\UI). 
    \label{eq:eigentaskproj}
\end{align}
The expected Capacity for the sensor sampled $S$ times to estimate $y^{(j)} (\UI)$ is given by $1/(1+\beta^2/S)$. In other words $S/\beta^2$ is the expected SNR of approximating that function with inputs sampled from the prior $p(\bm{u})$. The eigentasks and their associated SNRs are specific to the prior $p(\bm{u})$, the embedding and the remaining circuit parameters.
\\
This formulation provides a systematic way to identify and rank linear combinations of features according to their information content since eigentasks with large SNR (i.e., small  $\beta_j^2/S$ ) are more robust to sampling noise and therefore better suited for generalization. This observation suggests a dimensionality-reduction technique for supervised ML based on an estimator built from the first $K_L$ eigentasks. A practical application of this feature-reduction technique to a specific ML problem implemented on a quantum processor suggested a cutoff $K_L$ such that $ \beta_{K_L}^2 \sim S $, later analytically confirmed in Ref.~\cite{hu_generalization_2024}.

Consequently, the resulting estimator, based on the most informative eigentasks, takes the form
\begin{align}
    f(\bm{u}) = \sum_{j=1}^{K_L} w_{{\rm et},j} \, y^{(j)}(\bm{u}), 
    \label{eq:estimator_eigentasks}
\end{align}
where the weights $w_{{\rm et},j}$ are trained on the reduced feature set. Finally, although the definition of eigentasks so far has assumed access to the idealized, infinite-shot features $\bm{x}$ and an infinitely large training set, they can still be effectively approximated and remain practically useful under finite measurement shots and finite  training dataset size~\cite{PhysRevX.13.041020}.

\begin{figure}[t]
    \centering
    \includegraphics[scale=1.0]
    {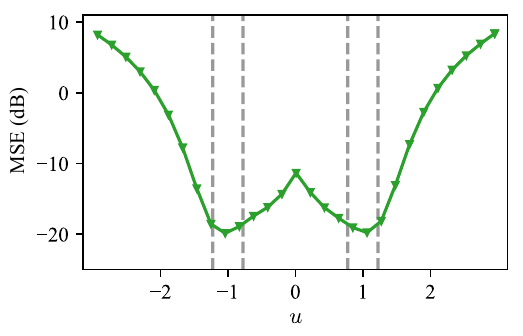}
    \caption{Results of the optimization for approximating the target function $f(u)=u$ where $u$ is sampled from a symmetric mixture of Gaussian prior distributions of width $\sigma^2=0.05$ and for a circuit with $n_{\rm en}=1$, $n_{\rm de}=0$, $L=32$, and $S=1$. Mean squared error, ${\rm MSE}(u)$, as a function of the true input phase $u$. The dashed grey lines correspond to the standard deviation around the respective means. The achieved precision is $\EUI{{\rm MSE}} \approx -18.3~{\rm dB}$ compared to the optimal allowed $\EUI{{\rm MSE}}_{\rm ult} \approx -22.5~{\rm dB}$.} 
    \label{fig:MSE_mixture_gaussians}
\end{figure}

\subsection{Results}
\label{sec:results}

In this section we apply the general framework developed above in approximating arbitrary functions of a single parameter $u$ which is drawn from a complex prior distribution. As already emphasized this is in stark contrast to more conventional Fisher information-based approaches which optimize accuracy only locally corresponding to $\sigma \rightarrow 0$ and are valid over a narrow dynamic range of the order $\sigma\sim 1/L$. Furthermore, following the sensing pipeline outlined in Sec.~\ref{sec:PNN_metrology}, we consider a training dataset consisting of $N_{\rm tr}=10000$ samples independently drawn from the corresponding probability distribution. Finally, for all the results presented in this section we fix $S=1$ i.e assuming single-shot experiments. While the choice $S=1$ may initially appear restrictive, it is directly relevant to sensing scenarios where the signal varies between repetitions but is sampled from the same underlying distribution. Moreover, analyzing performance in the single-shot regime is critical for future applications involving large quantum many-body systems. In such systems, the exponentially large number of possible measurement outcomes, combined with a finite number of experimental repetitions, implies that most outcomes occur at most once, or not at all, thereby effectively exhibiting the characteristics of the single-shot sensing regime in practice.

Motivated by the successful application of a variational quantum  algorithm in a trapped-ion platforms, and following Refs.~\cite{PhysRevX.11.041045,Marciniak2022}, we assume that the probe, entangler and decoder circuits take the form 
\begin{align}
U_{\rm em}(u) &= \mathcal{R}_z(u), \\
U_{\rm en}(\FSen) &= \mathcal{R}_x(\theta_{n_{\rm{en}}}^{(3)}) \mathcal{T}_x (\theta_{n_{\rm{en}}}^{(2)}) \mathcal{T}_z (\theta_{n_{\rm{en}}}^{(1)}) \cdots \notag \\
            &\quad \mathcal{R}_x(\theta_{1}^{(3)}) \mathcal{T}_x (\theta_{1}^{(2)}) \mathcal{T}_z (\theta_{1}^{(1)}) \mathcal{R}_y\left(\frac{\pi}{2}\right), \label{eq:U_en} \\
U_{\rm de}(\FSde) &= \mathcal{R}_x\left(\frac{\pi}{2}\right) \mathcal{T}_z (\vartheta_{1}^{(1)}) \mathcal{T}_x (\vartheta_{1}^{(2)}) \mathcal{R}_x (\vartheta_{1}^{(3)}) \cdots \notag \\
            &\quad \mathcal{T}_z (\vartheta_{n_{\rm{de}}}^{(1)}) \mathcal{T}_x (\vartheta_{n_{\rm{de}}}^{(2)}) \mathcal{R}_x (\vartheta_{n_{\rm{de}}}^{(3)}). \label{eq:U_de}
\end{align}
Here, $n_{\rm en}$ and $n_{\rm de}$ denote the number of layers in the entangler and decoder, respectively. The collective rotations and one-axis twisting operations are defined as $\mathcal{R}_{\alpha}(\phi) = e^{-i\phi J_{\alpha}}$ and $\mathcal{T}_{\alpha}(\phi) = e^{-i\phi J_{\alpha}^2}$, respectively, with $\alpha \in \{x, y, z\}$. The collective spin operators are given by $J_{\alpha} \equiv \sum_{i=1}^L \hat{\sigma}_{\alpha}^{(i)}/2$, where $\hat{\sigma}_{\alpha}^{(i)}$ denotes the Pauli-$\alpha$ operator acting on the $i$-th qubit. We assume a projective measurement of the $J_z$ observable, with eigenstates $\ket{j;m}$ satisfying $\bm{J}^2 \ket{j;m} = j(j+1)\ket{j;m}$ and $J_z \ket{j;m} = m\ket{j;m}$, where $j$ and $m$ correspond to the total angular momentum and its projection along the $z$-axis, respectively. Note that since our protocol employs global operations that conserve the total angular momentum, it is sufficient to restrict the analysis to the $j = L/2$ subspace, which is also the sector in which the maximum precision is attained.
$X_k(u)$, the feature we measured, is the probability of the $z$-component of the total angular momentum. In practice, this $z$-component is measured by performing an ordinary computational basis measurement, and $X_k = k/L$ where $k$ is the number of spin-up outcomes. Following the efficient optimization algorithm described in Sec.~\ref{sec:resevoir_optimization} and further analyzed in Appendix~\ref{sec:reservoir_direct}, we determine the optimal parameters $\FSen$, $\FSde$, and $\bm{w}$ that minimize the loss function in Eq.~\eqref{eq:loss}.

\begin{figure}[t]
    \centering
    \includegraphics[scale=1.0]
    {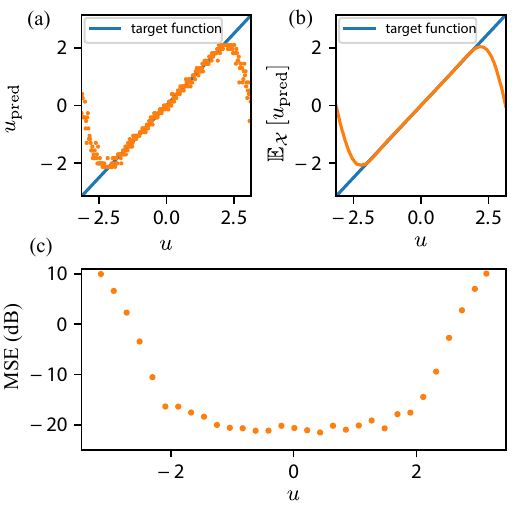}
    \caption{
Results of the optimization for $n_{\rm en}=1$, $n_{\rm de}=2$, $\sigma=0.7981$, $L=32$, and $S=1$. 
(a) Predicted output $u_{\rm pred}$ (orange dots) as a function of the input phase $u$ for a single experimental repetition. The target function $f^*(u) = u$ is shown in blue. Each dot corresponds to a distinct measurement outcome, highlighting the stochasticity of the single-shot regime. 
(b) Averaged prediction $\langle u_{\rm pred} \rangle$ over many repetitions, demonstrating convergence to the target function in a wide range around $u=0$. 
(c) Mean squared error ${\rm MSE}(u)$ as a function of the true phase $u$, quantifying the estimator's performance across the domain. Importantly, the achieved $\EUI{{\rm MSE}} \approx -19.1~{\rm dB}$ is remarkably close to the ultimate bound of $\EUI{{\rm MSE}}_{\rm ult} \approx -20~{\rm dB}$ allowed by quantum mechanics.
} 
    \label{fig:Figure_predictions}
\end{figure}
\textit{Approximating a non-linear function---}To illustrate the importance of our framework, we consider the task of approximating a non-trivial target function, $f(u) = \sin(10u)$, where the unknown parameter $u$ is drawn from a Gaussian prior distribution 
\begin{align}
    P_{\sigma}(u) = \frac{1}{\sqrt{2 \pi} \sigma} \exp\left(-\frac{u^2}{2 \sigma^2}\right),
    \label{eq:gaussian_prior}
\end{align}
while the domain is restricted to $u \in [-\pi,\pi]$ by rejecting samples outside this interval. We note that given the Gaussian form of the prior distribution, the training dataset size can be substantially reduced, if needed, by approximating the loss function using a Gauss–Hermite quadrature integration, as detailed in Appendix~\ref{sec:reservoir_direct}. 

As a first naive strategy, one could attempt to optimize the quantum circuit and the estimator to approximate the parameter $u$ itself, by constructing a linear estimator of the form $u_{\rm est} = \bm{w}^T \mathbf{X}(\UI)$. The target function is then approximated indirectly by applying $f$ to the estimated parameter, i.e., $f(u_{\rm est}) = \sin(10u_{\rm est})$. 

In contrast, within our approach, the entire sensing pipeline—including both the circuit and the estimator—is co-optimized directly for the task of approximating the target function $f(u) = \sin(10u)$, rather than the parameter $u$ itself. Specifically, we seek to construct a linear estimator of the form $f_{\rm est} = \bm{w}^T \mathbf{X}(\UI)$ that best approximates $f(u)$ by minimizing the cost function defined in Eq.~\eqref{eq:loss}.

\begin{figure}[t]
    \centering
    \includegraphics[scale=1.0]
    {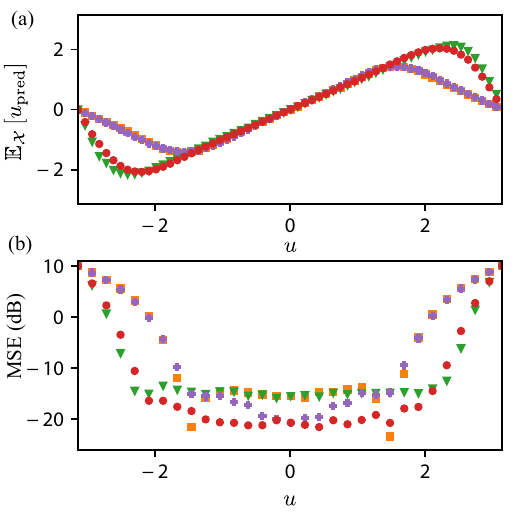}
    \caption{Results of the optimization for various circuit ans\"atze. 
        Orange squares correspond to a circuit with depth $n_{\rm en} = 0$ and $n_{\rm de} = 0$, 
        green triangles to $n_{\rm en} = 1$ and $n_{\rm de} = 0$, 
        purple crosses to $n_{\rm en} = 0$ and $n_{\rm de} = 2$, 
        and red dots to $n_{\rm en} = 1$ and $n_{\rm de} = 2$. 
        (a) Average predicted output as a function of $u$ for the different circuit depths. 
        (b) Corresponding mean squared error for the same circuit configurations.
} 
    \label{fig:various_circuits}
\end{figure}

This task-oriented optimization ensures that both the feature extraction performed by the circuit and the final estimator are tailored to the inference objective. As a result, it can yield significantly improved performance—especially when the target function is highly nonlinear, such as $f(u) = \sin(10u)$. This improvement is clearly illustrated in Fig.~\ref{fig:MSE_sin(10u)}, where we show the mean squared error ${\rm MSE}(u) \equiv \mathbb{E}\left[ \left( u - u_{\rm pred} \right)^2 | u \right]$, as function of the true value of the unknown phase for $\sigma=0.7981$ and using a circuit with $n_{\rm en} = 1$ and $n_{\rm de} = 0$ layers. 

To make the comparison even more concrete, in the naive two-step approach—where the circuit is first optimized to estimate $u$ and the nonlinearity is applied afterward—we achieve an average  error of $\EUI{{\rm MSE_1}} \approx 0.51 \approx -2.9~{\rm dB}  $. In contrast, using our direct optimization strategy, the error is significantly reduced to $\EUI{{\rm MSE_2}} \approx 0.34 \approx -4.7~{\rm dB}$, highlighting the practical advantage of our task-specific co-optimization.

\textit{Function approximation for signals with a non-trivial prior}---We next demonstrate our framework in the setting of estimating a single parameter $u$ drawn from a non-trivial prior distribution. Here, the prior should be understood as the predictive distribution of the signal immediately before interrogation, determined by the underlying signal dynamics and by any information available from previous measurements. Thus, although our framework does not impose a particular functional form on the prior, $P(u)$ is in general physically motivated rather than chosen arbitrarily. Such Bayesian priors arise naturally, for example, in phase and frequency estimation, atomic-clock operation, and the adaptive tracking of fluctuating signals \cite{demkowicz2011,bonato2017,Marciniak2022}. As a representative example illustrating the flexibility of the framework, we consider the bimodal Gaussian prior
\begin{align}
    P(u) = &\frac{c_1}{\sqrt{2 \pi} \sigma_1} \exp\left(-\frac{(u-\mu_1)^2}{2 \sigma_1^2}\right)+\\
    &\frac{c_2}{\sqrt{2 \pi} \sigma_2} \exp\left(-\frac{(u-\mu_2)^2}{2 \sigma_2^2}\right).
    \label{mixed_gaussian_prior}
\end{align}
We choose $c_1=c_2=1/2$, $\mu_1=1$, $\mu_2=-1$ and $\sigma_1^2=\sigma_2^2=0.05$. As illustrated in Fig.~\ref{fig:MSE_mixture_gaussians} where we employ a circuit with $n_{\rm en} = 1$ and $n_{\rm de} = 0$ layers, our approach achieves high estimation accuracy in regions near the peaks of the Gaussian components—i.e., within one standard deviation of each prior mean—while the performance naturally degrades in regions of low prior probability.

\begin{figure}[t]
    \centering
    \includegraphics[width=0.48\textwidth]
    {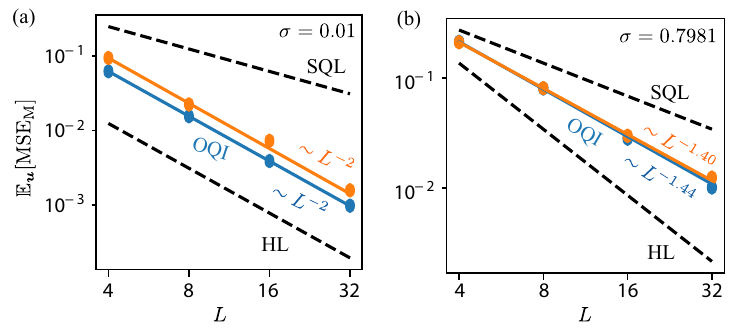}
    \caption{Scaling of the effective measurement variance $\EUI{{\rm MSE}_{\rm M}}$ as a function of the system size $L$ for $n_{\rm en}=1$, $n_{\rm de}=2$ circuit and for (a) a narrow prior with $\sigma = 0.01$ and (b) a wide prior with $\sigma = 0.7981$. The blue solid line shows the scaling predicted by the optimal quantum interferometer (OQI), while the orange solid line shows the scaling achieved by our optimization algorithm. The upper and lower black dashed lines indicate the standard quantum limit (SQL) and the Heisenberg limit (HL), respectively.} 
    \label{fig:scaling_L}
\end{figure}

\textit{Benchmarking}---Finally, and for benchmarking our framework with existing results in literature; see Refs.~\cite{PhysRevX.11.041045,Marciniak2022}, we apply our framework in estimating a single parameter $u$ drawn from the Gaussian prior of Eq.~\eqref{eq:gaussian_prior} with $\sigma=0.7981$. Results of the optimization for $n_{\rm en}=1$, $n_{\rm de}=2$, $\sigma=0.7981$, $L=32$, and $S=1$ are shown in Fig.~\ref{fig:Figure_predictions}. In particular, Fig.~\ref{fig:Figure_predictions}(a) illustrates the approximation of the target function $f^*(u) = u$ (shown in blue) based on a single experimental shot. Due to the stochastic nature of the measurement outcomes, different repetitions of the experiment yield different predictions. Averaging over many repetitions gives rise to the functional form shown in Fig.~\ref{fig:Figure_predictions}(b), which successfully captures the shape of the target function over a broad range around $u = 0$. The lower panel displays the mean squared error,
${\rm MSE}(u)$, as a function of the true value of the unknown phase. Importantly, even for such shallow circuit, the achieved average error is $\EUI{{\rm MSE}} \approx -19.1~{\rm dB}$, which is very close to the ultimate limit of an optimal quantum interferometer (OQI) of $\EUI{{\rm MSE}}_{\rm ult} \approx -20~{\rm dB}$ as imposed by quantum mechanics~\cite{Macieszczak_2014}.

We further investigate the performance of our algorithm using various circuit ans\"atze, as illustrated in Fig.~\ref{fig:various_circuits}, incorporating both tailored input states and effective POVMs. Our results are in complete agreement with those previously reported, thereby saturating the existing benchmarks in the literature and validating the effectiveness of our framework in reproducing state-of-the-art performance.

Going one step beyond we further study the scaling of an effective measurement variance $\EUI{{\rm MSE}_{\rm M}}$, where $1/\EUI{{\rm MSE}_{\rm M}} \equiv 1 / \EUI{{\rm MSE}}-1/\sigma^2$\cite{PhysRevX.11.041045,PhysRevResearch.6.023179}, with respect to the system size $L$ as illustrated in Fig.~\ref{fig:scaling_L} for $n_{\rm en}=1$, $n_{\rm de}=2$ and for two representative values of $\sigma$. In both cases our optimization algorithm attains the scaling of an OQI and goes beyond the scaling of the standard quantum limit (SQL) while for sufficiently narrow priors it reaches the Heisenberg limit (HL), as expected. 

\begin{figure}[t]
    \centering
    \includegraphics[scale=1.0]
    {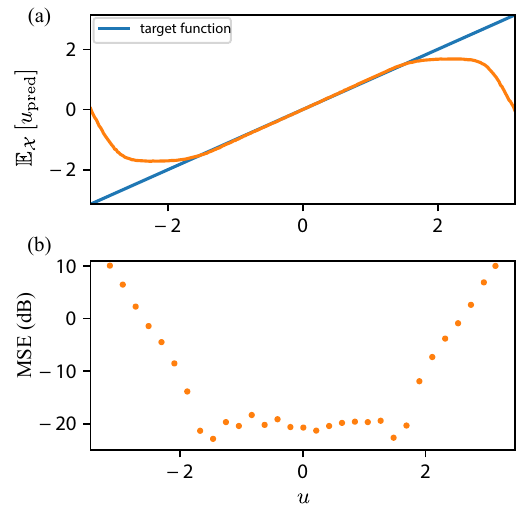}
    \caption{ Results of the optimization for $n_{\rm en}=1$, $n_{\rm de}=2$, $\sigma=0.7981$, $L=32$, and $S=1$ using the the first three non-trivial eigentasks with the highest SNR. (a) Averaged prediction $\langle u_{\rm pred} \rangle$ over many repetitions, demonstrating convergence to the target function in a wide range around $u=0$. 
(b) Mean squared error ${\rm MSE}(u)$ as a function of the true phase $u$, quantifying the estimator's performance across the domain. The achieved $\EUI{{\rm MSE}} \approx -17.1~{\rm dB}$, although degraded, is remarkably close to the one with the use of all features illustrated in Fig.~\ref{fig:Figure_predictions}(c). This suggests that incorporating additional eigentasks can further enhance the estimation performance.
 } 
    \label{fig:Figure_predictions_eigentasks}
\end{figure}

\subsubsection{Use of eigentasks}
\label{sec:use_of_eigentasks}
So far, the optimization of our sensing pipeline has relied on the full set of raw features obtained from the sensor to successfully approximate the target function. However, as discussed in Sec.~\ref{sec:eigentasks}, a promising alternative is to consider linear combinations of features—referred to as \emph{eigentasks}—that exhibit a sufficiently high signal-to-noise ratio. Consequently, the optimization of the internal parameters of the entangling and decoding part of the circuit ansatz, as well as the estimator, must be revisited, since in principle the resulting optimal configurations generally depend on the number of eigentasks retained. Such modification arises from the cutoff criterion used to select eigentasks based on their signal-to-noise ratio, $\beta_k^2$. In the special case where all eigentasks are preserved, the optimization reduce to the original setting, due to the linearity of both the feature transformation and the estimator.

\begin{figure}[t]
    \centering
    \includegraphics[scale=1.0]
    {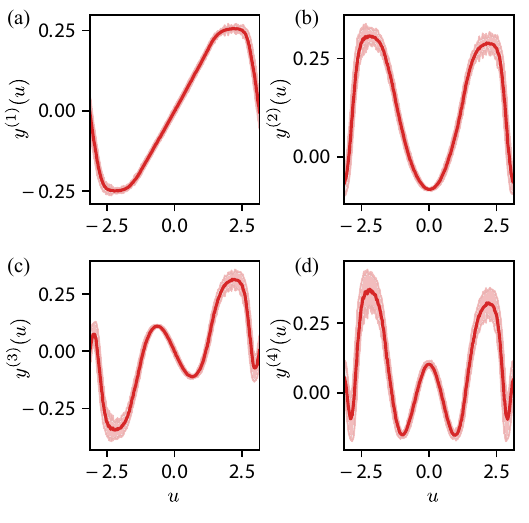}
    \caption{ Functional form of the mean values (solid lines) and standard deviations (shaded regions)  of the first four non-trivial eigentasks, shown in panels (a)-(d) respectively. These are obtained for the circuit ansatz and learning task described in Sec.~\ref{sec:use_of_eigentasks}. Notably, the first eigentask, panel (a), already reproduces the structure of the target function, $f(u)=u$, in a wide range around $u=0$. As expected, the uncertainty in the mean values increases with the eigentask index (see main text). 
} 
    \label{fig:eigentasks}
\end{figure}

Results in the special case of $n_{\rm en}=1$, $n_{\rm de}=2$, $L=32$ and $S=1$ are shown in Fig.~\ref{fig:Figure_predictions_eigentasks}, where the sensing pipeline has been optimized based on the three eigentasks with the highest SNR as defined in Eq.~\eqref{eq:eigenprob} and Eq.~\eqref{eq:eigentaskproj}. We highlight that the eigentask coefficients defined in Eq.~\eqref{eq:eigenprob} are computed in the infinite-shot regime, but based on a finite training dataset of size $N_{\rm tr} = 10000$. Surprisingly, we observe that the performance achieved using only a small number of eigentasks, although degraded, closely matches that of the full-feature case shown in Fig.~\ref{fig:Figure_predictions}, while the optimal parameters of the entangling and decoding circuits remain unchanged. Moreover, as shown in Fig.~\ref{fig:eigentasks}(a), the first non-trivial eigentask alone essentially reproduces the functional form of the target function, highlighting its dominant contribution.

\subsubsection{Adaptation to device imperfections and persistence of the quantum computational-sensing advantage}
\label{sec:robustness_noise}

Finite-shot uncertainty and device imperfections enter the inference problem in fundamentally different ways. Finite-shot uncertainty is the intrinsic sampling fluctuation of the empirical measurement record $\mathbf{X}^{(S)}(u)$ about the implemented outcome-probability vector $\mathbf{x}_{\rm e}(u)$; for an ideal device, $\mathbf{x}_{\rm e}(u)=\mathbf{x}(u)$ is determined by the Born rule. This irreducible sampling uncertainty is already incorporated into our loss function. Device imperfections instead alter the implemented input-output map itself through mechanisms such as measurement errors, decoherence, coherent miscalibration, or temporal drift. Stationary distortions that remain fixed between training and inference can be incorporated directly into the learned hardware response. We therefore examine two complementary cases: readout errors, modeled as a classical channel acting on the outcome distribution, and static coherent gate disorder, which modifies the quantum dynamics and breaks collective symmetry examined in the manuscript. These examples test whether end-to-end optimization can adapt the circuit and estimator to the realized device map and whether the computational-sensing advantage persists under such imperfections.

We test the extent of the expected hardware-aware adaptation using the linear target $f^{*}(u)=u$ as a metrological benchmark. In each case, the imperfection is stationary and incorporated consistently during training and inference, so that the optimization acts on the implemented hardware response rather than an ideal surrogate. For readout errors, we evaluate inference under the distorted outcome distribution and test whether the computational-sensing advantage for direct inference of $f(u)=\sin(10u)$ persists. For static coherent gate disorder, we compare full reoptimization of the circuit and estimator with a baseline that retains the circuit optimized for the ideal device and retrains only the classical estimator. These comparisons isolate the benefit of adapting the complete sensing pipeline to the realized device map.

We first consider imperfect readout. Since the measurement performed in the collective-spin setting is the total magnetization, we model readout noise as a classical noisy channel acting on the ideal magnetization distribution. If $P_m(u)$ denotes the ideal probability of obtaining magnetization outcome $m$ for an input parameter $u$, the experimentally observed distribution is modeled as

\begin{equation}
\widetilde{P}_{m}(u) = \sum_{m'} \Gamma_{m,m'} P_{m'}(u),
\label{eq:noisy_readout_channel_main}
\end{equation}
where $\Gamma_{m,m'}$ is the conditional probability that the ideal outcome $m'$ is recorded as $m$. Following Refs.~\cite{PhysRevA.98.030303,25ds-9724}, we use the effective Gaussian readout kernel

\begin{equation}
\Gamma_{m,m'}=\frac{\exp\!\left[-(m-m')^2/(2\sigma_{\rm ro}^2)\right]}{\sum_{m''}\exp\!\left[-(m''-m')^2/(2\sigma_{\rm ro}^2)\right]
},
\label{eq:gaussian_readout_kernel_main}
\end{equation}
where $\sigma_{\rm ro}$ controls the readout resolution. This model captures the finite resolution associated with realistic collective-spin measurements, for example due to fluorescence collection, thresholding, or coarse-grained magnetization detection. A microscopic derivation of this effective kernel from independent single-qubit assignment errors is provided in Appendix~\ref{app:readout_errors}. In particular, for symmetric assignment errors with probability $p$, the coarse-grained magnetization readout approaches a Gaussian kernel at large $L$, with a mean attenuated by a factor $(1-2p)$ and variance proportional to $Lp(1-p)$.

\begin{figure}[t!]
\centering{}
\includegraphics[width=0.48\textwidth]{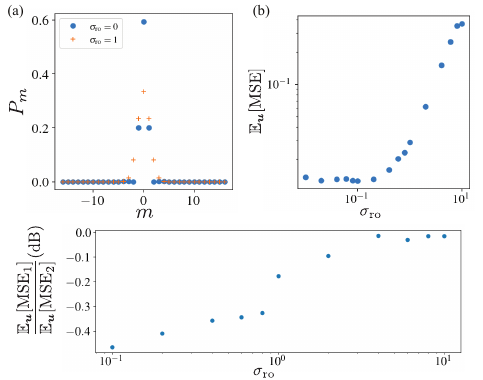} 
\caption{Effect of readout noise on measurement statistics and inference performance for $n_{\rm en}=1$, $n_{\rm de}=2$, $\sigma=0.7981$, $L=32$, and $S=1$. (a) Probability distribution of the measured magnetization outcomes, $m$, for $u=0$, shown for two representative values of the readout-noise strength $\sigma_{\rm ro}$. Increasing $\sigma_{\rm ro}$ broadens the observed distribution
thereby reducing the distinguishability of the measurement outcomes. (b) Inference performance of estimating the linear target $f^{*}(u)=u$ as a function of the readout-noise strength $\sigma_{\rm ro}$. The noisy readout channel is included consistently during both training and inference, allowing the end-to-end optimization to adapt the parametrized quantum circuit and classical estimator to the modified measurement statistics. The optimized protocol retains high inference performance over a broad range of readout-noise strengths. (c) Persistence of the advantage of the computational sensing approach, $\EUI{{\rm MSE_2}}$, estimating $f(u)=\sin(10u)$ over the indirect approximation where only the final classical layer is optimized, $\EUI{\rm MSE_1}$, as a function of the readout-noise strength $\sigma_{\rm ro}$}.
\label{fig:readout_noise} 
\end{figure}

Importantly, the noisy probabilities $\widetilde{P}_{m}(u)$ are used consistently both during training and testing allowing direct optimization of the circuit and estimator in the presence of the noisy readout channel. As shown in Fig.~\ref{fig:readout_noise}(b), the optimized protocols remain robust over a broad range of readout resolutions. 
The degradation in performance is gradual, indicating that the end-to-end procedure can identify feature combinations that remain informative even when the magnetization distribution is broadened by measurement noise; see Fig.~\ref{fig:readout_noise}(a). Finally, from Fig.~\ref{fig:readout_noise}(c) it is clear that the advantage of computational sensing approach on approaching functions such as $f(u)= \sin({10u})$, persists in a wide range of readout noise over the alternative of indirect approximation where only the final classical layer is optimized.

We next consider coherent gate imperfections that break the ideal collective-spin symmetry. This is a qualitatively different error source from readout noise. While readout noise can be incorporated as a classical channel acting on the measured probability distribution, gate errors modify the quantum dynamics itself and can drive the state outside the symmetric Dicke manifold. Consequently, this analysis requires simulations in the full $2^L$-dimensional Hilbert space rather than in the reduced symmetric subspace. This substantially increases the numerical cost, and we therefore restrict this part of the analysis to a representative system size $L=4$, with $n_{\rm en}=1$, $n_{\rm de}=2$, and prior width $\sigma_u=0.7981$.

As a reference point, we first perform the optimization in the full Hilbert space without adding gate disorder. This allows us to verify that indeed the end-to-end optimization naturally selects solutions close to the symmetric Dicke manifold even when the symmetry constraint is not explicitly imposed. 

\begin{figure}[t!]
\centering{}
\includegraphics[width=0.48\textwidth]{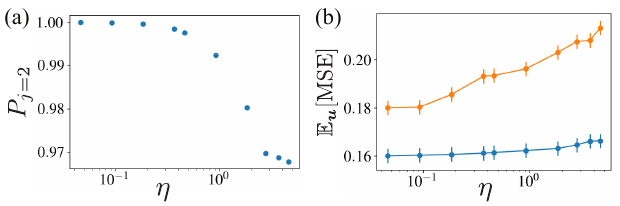} 
\caption{Effect of gate imperfections on dynamics and inference performance for $n_{\rm en}=1$, $n_{\rm de}=2$, $\sigma=0.7981$, $L=4$, and $S=1$. (a) Probability of remaining in the maximal symmetric subspace of $j=2$ for $u=0$, as a function of noise $\eta$. (b) Inference performance of estimating the linear target $f^{*}(u)=u$ as a function of the strength of gate disorder. End-to-end optimization (blue dots) remains robust over a broad range of disorder strengths and consistently outperforms the case in which the internal circuit parameters are fixed to their noiseless optimal values and only the classical output layer is retrained (orange dots). }  
\label{fig:gate_imperfections} 
\end{figure}

We then introduce symmetry-breaking gate imperfections in the one-axis-twisting operations. We assume that the disorder realization is fixed throughout training, corresponding to static calibration errors or spatial inhomogeneities rather than iteration-to-iteration drift. In the ideal collective model, one-axis twisting is generated by collective quadratic spin operators. To model imperfect all-to-all interactions, we replace the ideal collective twisting Hamiltonian by a disordered two-body Hamiltonian,

\begin{equation}
\mathcal{T}_{\alpha}(\phi)=\exp\!\left[-i\phi H_{\alpha}^{\rm ge}\right],
\label{eq:gate_error_twist}
\end{equation}
with $H_{\alpha}^{\rm ge}=\sum_{i<j} J_{ij}^{(\alpha)}
\hat{\sigma}_{\alpha}^{(i)} \hat{\sigma}_{\alpha}^{(j)}$,
and $J_{ij}^{(\alpha)} = J+\delta J_{ij}^{(\alpha)}$.
Here $\alpha\in\{x,y,z\}$ labels the twisting axis, and the coupling fluctuations are sampled independently from a normal distribution,
\begin{equation}
\delta J_{ij}^{(\alpha)}
\sim
\mathcal{N}(0,\eta^2),
\label{eq:gate_error_distribution}
\end{equation}
where $\eta$ quantifies the strength of the gate disorder. Physically, this model captures calibration errors, spatial inhomogeneities, and imperfect control fields that break the permutation symmetry of the ideal collective interaction. Such imperfections generate leakage outside the symmetric subspace, as illustrated in Fig.~\ref{fig:gate_imperfections}(a),  and therefore test whether the optimized sensing protocol relies sensitively on exact collective symmetry.

The results are shown in Fig.~\ref{fig:gate_imperfections}. Although the disordered twisting gates induce nonsymmetric components in the wavefunction, the end-to-end optimized protocol remains robust. In particular, the optimization partially compensates for the symmetry-breaking dynamics by readjusting the circuit parameters and the estimator weights, thereby redistributing population across Hamming-weight sectors in a way that preserves task performance and naturally outperforms the case in which the internal circuit parameters are fixed to their noiseless optimal values and only the classical output layer is retrained. This demonstrates that the optimized protocol is not simply exploiting a fine-tuned ideal collective trajectory, but can adapt to coherent imperfections that perturb the accessible quantum features.

\subsubsection{Extension to arbitrary tasks }
\label{sec:arbitrary_tasks}

\begin{figure}[t!]
\centering{}
\includegraphics[width=0.48\textwidth]{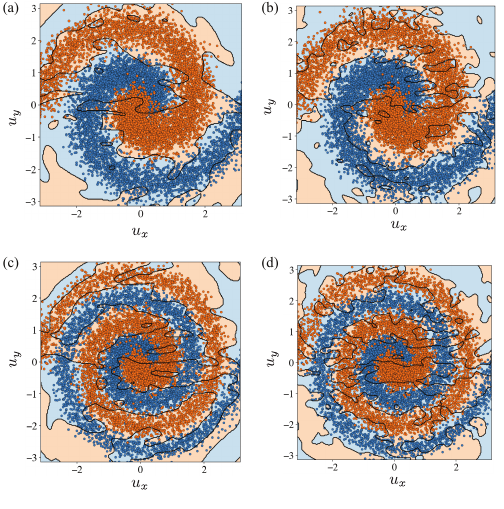} 
\caption{Binary classification of the noisy ``two spirals'' dataset for $L=32$, $n_{\rm en}=1$ and $n_{\rm de}=2$ with infinite shots. The background colors indicate the predicted class, revealing the learned decision boundaries. Panels (a,c) correspond to the direct (end-to-end, resource-aware) optimization approach, while panels (b,d) use the indirect approximation strategy. The top row (a,b) shows the one-turn spiral dataset ($S1$), and the bottom row (c,d) the more challenging two-turn case ($S2$). 
The direct approach achieves classification accuracy of $0.84$ and $0.74$ for $S1$ and $S2$, respectively, compared to $0.75$ and $0.70$ obtained with the indirect method and the same training resources. Note that the illustrated testing dataset is approximately ten times larger than the training one.}
\label{fig:spiral_infinite_shots_noise} 
\end{figure}

So far, we have focused on function-approximation tasks arising in regression and parameter estimation. In this section, we demonstrate that the proposed end-to-end optimization framework is not restricted to such settings, but can also be applied to genuinely nonlinear supervised-learning problems. As a representative benchmark, we consider binary classification on the standard two-spirals dataset \cite{chen2023}. This dataset is a prototypical example of a nonlinearly separable classification problem: the two classes form intertwined manifolds in the input space, so that successful classification requires learning a nontrivial decision boundary rather than approximating a single one-dimensional target function.

At first, we use the same variational architecture as in the previous sections. The only modification is that the scalar input is replaced by a two-dimensional input
$\mathbf{u}=(u_x,u_y)$, which is embedded through the collective-spin unitary

\begin{equation}
U_{\rm em}(\bm{u})=\exp\left[-i\left(u_x J_x+u_y J_y\right)\right].
\label{eq:emb_spiral}
\end{equation}
 After this embedding, the state is processed by the same structure of entangling and decoding layers used throughout the manuscript, and the measurement outcomes are used as features for a classical classifier. We compare two optimization strategies. In the direct approach, the quantum circuit is optimized end-to-end for the final classification task, including the effect of the available measurement resources. In the indirect approach, the circuit is instead optimized according to an intermediate approximation objective of minimize bayesian risk in estimating $(u_x,u_y)$, after which the resulting features are used for classification.

The results are shown in Fig.~\ref{fig:spiral_infinite_shots_noise} for a system with $L=32$, $n_{\rm en}=1$, and $n_{\rm de}=2$. The background colors indicate the predicted class and therefore visualize the learned decision boundary. We consider both a one-turn spiral dataset ($S1$) and a more challenging two-turn spiral dataset ($S2$). In the infinite-shot limit shown in Fig.~\ref{fig:spiral_infinite_shots_noise} , the direct end-to-end approach achieves an accuracy of $0.84$ for the one-turn spiral, compared to $0.75$ for the indirect strategy. For the two-turn spiral, where the two classes are more strongly intertwined, the direct approach achieves an accuracy of $0.74$, compared to $0.70$ for the indirect strategy. Thus, for the two spiral datasets and circuit architecture considered here, direct end-to-end optimization yields higher infinite-shot classification accuracy than the indirect strategy, with gains of $9$ and $4$ percentage points for $S1$ and $S2$, respectively. Similar trends persist in the finite-shot regime (e.g., $S = 100$ or $S = 1$), where the task becomes significantly more challenging and resource-awareness plays a critical role. In particular, we notice that $S = 100$ already suffice to saturate the performance of infinite shots while for $S = 1$ and for the one-turn
spiral dataset we achieve an accuracy of $0.62$ and $0.54$ for the direct and indirect approach respectively.  We emphasize that in this regime the circuit is optimized under the same sampling constraints since the alternative of first optimizing an ideal infinite-shot representation and only subsequently evaluating it with finite measurement samples leads to significantly inferior performance as expected from our analysis.

One thing immediately apparent from Fig.~\ref{fig:spiral_infinite_shots_noise} is that the learned decision boundaries do not closely follow the underlying spiral manifolds, indicating limited expressivity of the present model. This mismatch is expected to constrain out-of-sample generalization, particularly for more intricate class geometries. The limitation can be traced to the restricted Fourier spectrum generated by the collective embedding, as analyzed in Appendix~\ref{app:arbitrary_tasks}.

This observation motivated us to increase the expressivity of the circuit ansatz by enriching the input embedding. To this end, we replace the permutation-symmetric collective embedding by the trainable local embedding

\begin{equation}
\label{eq:emb_spiral2}
U_{\rm em}(\bm{u}) = e^{-i \left( u_x \sum_{i=1}^{L}   h_x^{(i)}\sigma_x^{(i)} + u_y \sum_{i=1}^{L}h_y^{(i)}\sigma_y^{(i)} \right)}
\end{equation}
Here, $h_x^{(i)}$ and $h_y^{(i)}$ are trainable local embedding coefficients. Allowing different qubits to couple differently to the input coordinates breaks permutation symmetry and enlarges the family of input-dependent features accessible to the classifier. Because this embedding no longer preserves the symmetric Dicke subspace, the dynamics are simulated in the full $2^L$-dimensional Hilbert space, and the measured features are taken to be the probabilities of the individual computational-basis bitstrings. This differs from the permutation-symmetric model, for which computational-basis outcomes with the same Hamming weight are equivalent and the measurement is described by the $L+1$ probabilities associated with the possible $J_z$ outcomes. To keep the number of measured features comparable to the $L=32$ symmetric model, which yields $L+1=33$ such outcome probabilities, we perform the local-embedding simulations with $L=5$, yielding $2^5=32$ bitstring probabilities. The local embedding therefore provides a richer input-dependent feature map at essentially fixed feature dimension, increasing the model's capacity to represent the nontrivial decision boundaries required for the spiral-classification task. 

\begin{figure}[t!]
\centering{}
\includegraphics[width=0.48\textwidth]{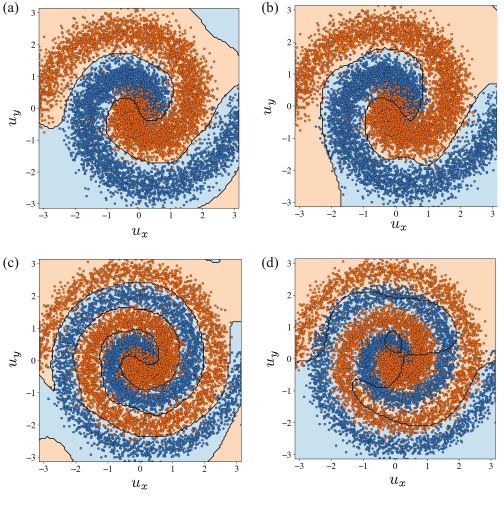} 
\caption{Binary classification of the noisy``two spirals'' dataset for the local embedding of Eq.\eqref{eq:emb_spiral2} with $L=5$, $n_{\rm en}=1$ and $n_{\rm de}=2$ with infinite shots. The background colors indicate the predicted class, revealing the learned decision boundaries. Panels (a,c) correspond to the direct (end-to-end, resource-aware) optimization approach, while panels (b,d) use the indirect approximation strategy. The top row (a,b) shows the one-turn spiral dataset, and the bottom row (c,d) the more challenging two-turn case. 
The direct approach achieves classification accuracy of $0.96$ and $0.92$ for the one- and two-turn datasets, compared to $0.95$ and $0.65$ obtained with the indirect method respectively. Note that the illustrated testing dataset is approximately ten times larger than the training one.} 
\label{fig:spiral_infinite_shots2} 
\end{figure}

The resulting decision boundaries are presented in Fig.~\ref{fig:spiral_infinite_shots2}. Compared with the original collective embedding, the modified ansatz produces a clearly more faithful separation of the intertwined spiral manifolds with accuracy $0.96$ and $0.92$ for the one-turn and two-turn dataset respectively and in the computational sensing approach. In the indirect approach we respectively achieve an accuracy of $0.95$ and $0.65$. We highlight that the reduced accuracy of the indirect approach on the two-turn dataset should not be interpreted as a failure of the circuit to encode the input coordinates. Indeed, the pairs $(u_x,u_y)$ can be estimated with very small error. Rather, the limitation arises in the subsequent post-processing step: a linear map from the reconstructed coordinates, or from the corresponding features, is not sufficiently expressive to implement the highly nonlinear decision rule required by the two-turn spiral. Resultingly, one would expect a better out-of-sample generalization capacity for the local ansatz. 

\section{Discussion and Conclusion}
\label{sec:conclusions}

We have developed a data-driven supervised learning framework for quantum sensing in which state preparation, encoding, measurement, and classical estimation are optimized jointly with respect to a global inference task. Formulating quantum sensing in supervised-learning terms allows direct approximation of arbitrary nonlinear functions of stochastic inputs, extending the scope of quantum metrology beyond local Fisher-information bounds and connecting performance to a broader notion of functional capacity.

Technically, we introduced an efficient optimization procedure that integrates a reservoir-inspired estimator with post-reservoir loss minimization, enabling convergence to near-optimal solutions even in the single-shot regime where sampling noise is strongest. The eigentask analysis further identifies noise-robust feature combinations that produce compact estimators while suppressing unstable or low-SNR modes.

Our results show that co-optimizing the entire sensing pipeline yields substantial gains over two-stage strategies in which parameter estimation precedes functional evaluation. The framework reproduces known benchmarks, accommodates general priors and nonlinear targets, and provides a unified approach for realistic sensing scenarios demonstrating robustness against various relevant noise sources. We further demonstrate that the same principle with similar efficiency enhancement naturally extends beyond one-dimensional function approximation where more complicated loss functions need to considered. In particular, by applying the framework to binary classification of the noisy two-spirals dataset, we show that the method can address genuinely nonlinear decision-making tasks requiring nontrivial decision boundaries. These findings suggest a shift in quantum metrology toward integrated, data-driven co-design of sensors and inference algorithms.

At the same time we highlight that the present implementation has important scalability considerations. The use of collective spin operations makes the number of variational parameters independent of the system size $L$, and the restriction to the symmetric subspace enables simulations at relatively large effective spin sizes. This is well motivated for the metrological tasks considered here, since optimal precision for estimating one-dimensional linear functions of a collective phase is known to be attainable within the maximum angular momentum sector allowing for high performance with shallow circuits. The situation is different for more general learning tasks, where no analogous optimality guarantee exists. Capturing more complex decision boundaries or higher-dimensional structure may require deeper circuits, local controls, or more expressive ans\"atze. In such cases, the dimensionality of the parameter space can increase substantially, and the DIRECT global-optimization routine used in the present work is not expected to remain the most efficient choice. A natural and common route toward this regime is to combine end-to-end optimization with continuation or warm-start strategies since in practice, we observe that optimized parameters for a given system size and circuit depth often provide good initializations for nearby configurations \cite{PhysRevX.10.021067, Skolik_2021, Egger_2021,PRXQuantum.6.010317}. Such strategies can be combined with gradient-based optimization methods, using either exact gradients when available or numerical approximations, and are expected to scale more favorably with the number of parameters than global search methods.

Several directions follow from this work. A deeper theoretical link between functional capacity and quantum information bounds in multi-parameter and multi-shot settings would broaden the conceptual foundation. On the practical side, adapting the optimization framework to noisy intermediate-scale devices, incorporating hardware-aware ansätze, and benchmarking on experimental platforms such as trapped ions and superconducting circuits are natural next steps. Extending the current linear-estimator architecture to nonlinear or adaptive readouts may unlock further opportunities for quantum-enhanced sensing and learning.

\section{ACKNOWLEDGMENTS}
\label{sec:acknowledgements}
We acknowledge Peter Zoller and Denis Vasilyev for fruitful discussions as well as Charles Fromonteil for critical reading and feedback on our manuscript. Research was sponsored by AFOSR MURI Grant No. FA9550-22-1-0203 and by the Army Research Office under Grant No. W911NF-25-1-0261. The views and conclusions contained in this document are those of the authors and should not be interpreted as representing the official policies, either expressed or implied, of the Army Research Office or the U.S. Government. The U.S. Government is authorized to reproduce and distribute reprints for Government purposes notwithstanding any copyright notation herein.

\bibliography{refs}

\clearpage

\appendix

\begin{widetext}

\section{Sensing Loss, Fisher Information and Bhattacharyya Information}
\label{app:Fisher_Bhattacharyya}

In this section we establish an explicit relation between the functional $L [f ; S]$ and \textit{Fisher information} $\mathcal{I} (u)$. To this end, we emphasize that $L[f;S]$ depends non-trivially on the sample size $S$ but is independent of the parameter $u$, whereas $\mathcal{I}(u)$ depends on $u$ and it exhibits a trivial linear scaling for $S$ independent shots.  It is thus reasonable to fix the value of $S$ and $u$ with the natural choice being
\begin{equation}
    S = 1, \quad \mathrm{and} \quad u = 0.
\end{equation}
Let us then suppose that the input $u$ obeys a \textit{zero-mean} Gaussian distribution:
\begin{equation}
    p (u) = \frac{1}{\sqrt{2 \pi} \sigma} e^{- \frac{u^2}{2 \sigma^2}} .
\end{equation}
An important lemma in probability theory is the Gaussian \textit{integration by parts} formula: for any function $F (u)$
\begin{equation}
    \int^{\infty}_{- \infty} F' (u) \frac{1}{\sqrt{2 \pi} \sigma} e^{- \frac{u^2}{2 \sigma^2}} \dd u = - \int^{\infty}_{- \infty} F (u) \frac{\dd}{\dd u} \left( \frac{1}{\sqrt{2 \pi} \sigma} e^{- \frac{u^2}{2 \sigma^2}} \right) \dd u.
\end{equation}
Namely,
\begin{equation}
    \mathbb{E}_u [F' (u)] = \frac{1}{\sigma^2} \mathbb{E}_u [u F (u)] .
\end{equation}
By plugging the functions $x_j (u) = \mathrm{Tr} \{ \hat{\rho} (u) \hat{M}_j \}$ into $F (u)$, we get \cite{PhysRevX.13.041020}
\begin{align}
    L [f=u ; S = 1] & = \mathbb{E}_u[f(u)^2] - \mathbb{E}_u [f (u) \bm{x} (u)]^T \left( \mathbf{G}+ \frac{1}{S} \mathbf{V} \right)^{- 1} \mathbb{E}_u [f (u) \bm{x} (u)] = \sigma^2 - \mathbb{E}_u [u \bm{x} (u)]^T \mathbf{D}^{- 1} \mathbb{E}_u [u \bm{x} (u)] \nonumber\\
    & = \sigma^2 - \sum_{j = 1}^K \mathbf{D}^{- 1}_{jj} \cdot \mathbb{E}_u [u x_j (u)]^2 = \sigma^2 - \sum_{k = 1}^K \frac{1}{\mathbb{E}_u [x_j (u)]} \cdot \left( \sigma^2 \mathbb{E}_u \left[ \frac{\dd x_j}{\dd u} (u) \right] \right)^2 \nonumber\\
    & = \sigma^2 - \sigma^4 \sum_{j = 1}^K \frac{1}{\mathbb{E}_u [x_j (u)]} \left( \mathbb{E}_u \left[ \frac{\dd x_j}{\dd u} (u) \right] \right)^2,
    \label{eq:single_shot_capacity}
\end{align}
where $\gr \equiv \EUI{\bm{x} \bm{x}^T}$, $\ci \equiv \EUI{\bm{\Sigma}}$ and $\mathbf{D}_{j,j'}=\delta_{j,j'} \EUI{x_j}$.
Therefore,
\begin{equation}
    \lim_{\sigma^2 \rightarrow 0} \frac{1}{\sigma^2} \left( 1 - \frac{1}{\sigma^2} L [f = u ; S = 1] \right) = \sum_{j = 1}^K \frac{1}{x_j (0)} \left( \frac{\dd x_j}{\dd u} (0) \right)^2 =\mathcal{I} (0) .
\end{equation}
In general, letting $u = \sigma v$ we have
\begin{equation}
    \mathbb{E}_u [F (u)] =\mathbb{E}_u \left[ F (0) + \sigma v F' (0) + \frac{1}{2} \sigma^2 v^2 F'' (0) + \cdots \right] = F (0) + \frac{\sigma^2}{2} F'' (0) + O (\sigma^4) .
\end{equation}
Therefore, from Eq.~\eqref{eq:single_shot_capacity} we can express $C [f = u ; S = 1]$ as
\begin{align}
    L [f = u ; S = 1] &= \sigma^2 - \sigma^4 \sum_{j = 1}^K \frac{1}{x_j (0) + \frac{\sigma^2}{2} x_j'' (0) + O (\sigma^4)} \left( x'_j (0) + \frac{\sigma^2}{2} x_j''' (0) + O (\sigma^4) \right)^2 \\
    &= \sigma^2 - \sigma^4  \underbrace{\sum_{j = 1}^K \frac{x'_j (0)^2}{x_j (0)}}_{\text{Fisher Information}} - \sigma^6  \underbrace{\sum_{j = 1}^K \left( \frac{x'_j (0) x_j''' (0)}{x_j (0)} - \frac{x'_j (0)^2 x_j'' (0)}{2 x^2_j (0)} \right)}_{\text{Bhattacharyya Information}}+\mathcal{O} (\sigma^8) , 
\end{align}
where higher order terms in $\sigma^2$-expansion give rise to \textit{Bhattacharyya Information}.

In other words, the \textit{single-shot sensing loss} serves as a generating functional for the Fisher information and, more generally, for the entire hierarchy of Bhattacharyya informations. The Fisher information quantifies the attainable accuracy of parameter estimation in the regime where the unknown parameter is restricted to a sufficiently local region. However, when the parameter is distributed more broadly, the Fisher information alone no longer provides a complete characterization of the fundamental estimation limits. In this regime, higher-order Bhattacharyya information becomes essential, as it offers a refined description of the ultimate precision bounds of the metrological system \cite{vasilyev_optimal_2024}.

There is a simple way to compare the derivation above and the Bhattacharyya bounds derivation in Ref.\,\cite{vasilyev_optimal_2024}. 
Supposing a quadratic loss for outcome $j$, we have
\begin{equation}
    L = \sum_{j = 1}^K \int (u - v_j)^2 x_j (u) p (u) \dd u,
\end{equation}
where $v_j$ can be regarded as a single-shot estimator when obtaining the outcome $j$. The optimal $\{ v_j \}$ is obtained by solving
\begin{equation}
    \frac{\partial L}{\partial v_j} = - 2 \int (u - v^{\ast}_j) x_j (u) p (u) \dd u = 0,
\end{equation}
whose solution is
\begin{equation}
    v^{\ast}_j = \frac{\int u x_j (u) p (u) \dd u}{\int x_j (u) p (u) \dd u} = \frac{\mathbb{E}_u [u x_j]}{\mathbb{E}_u [x_j]} .
\end{equation}
This is exactly the Eq.(C12) in Ref.\,\cite{PhysRevX.13.041020}, when taking $S = 1$ and $f (u) = u$; it is also the Eq.(14) in Ref.\,\cite{vasilyev_optimal_2024}.

\section{Cumulant Analysis of Function of Features with Quantum Shot Noise}
\label{appsec:cumulant-qsn}

In this section, we show an important result to characterize the behavior of any function of features with quantum shot noise. In Appendix \ref{appsec:post-rc-loss}, we will use this result to prove that the post-reservoir loss in Eq.\,(\ref{eq:post_reservoir_loss}) is approximately Gaussian.

We recall in Sec.\,\ref{sec:PNN_metrology} we define the noisy features as the empirical means of finite shots
\begin{equation}
    \bm{X} = \frac{1}{S} \sum_s \bm{X}^{(s)} = \bm{x} + \bm{\zeta},
\end{equation}
where all single-shot features $\bm{X}^{(s)}$ are i.i.d. sampled from physical systems. Here, we absorb the scaling factor $1/\sqrt{S}$ into the shot noise term $\bm{\zeta}$. According to the homogeneity of cumulants, we can let the $j$-th order cumulant of $\bm{\zeta}$ be $\kappa_j [\bm{\zeta}] = \frac{1}{S^{j - 1}} \bm{\kappa}^{(j)}$ where and $\bm{\kappa}^{(j)}$ must be an $S$-independent order-$j$ tensor. For example, $\kappa_1 [\bm{\zeta}] =\mathbb{E} [\bm{\zeta}] = \bm{0}$ and $\kappa_2 [\bm{\zeta}] = \mathrm{Cov} [\bm{\zeta}] = \frac{1}{S} \bm{\Sigma}$. Particularly for quantum sampling noise, we have $\bm{\kappa}^{(2)} = \bm{\Sigma} = \mathrm{diag} (\bm{x}) - \bm{x} \bm{x}^T$. In this section, we focus on the expectation over shot noise, conditioned on a fixed input $\UI$. Therefore, we use $\mathbb{E}[g]$ and $\mathrm{Cov}[g]$ to abbreviate $\mathbb{E}[g | \UI]$ and $\mathrm{Cov}[g | \UI]$ for arbitrary $g$. 

In this appendix, we are going to analyze the cumulants of a function of the noisy features $\bm{X} (\bm{u})$, and prove that $\mathcal{F} (\bm{X})$ can be approximated by

\begin{equation}
    \mathcal{F} (\bm{X}) \approx\mathcal{N} \left( \mathcal{F} (\bm{x}) + \frac{1}{2 S} \sum_{j, j'} \frac{\partial^2 \mathcal{F} (\bm{x})}{\partial z_k \partial z_{j'}} \Sigma_{j j'}, \frac{1}{S} \sum_{j, j'} \left( \frac{\partial \mathcal{F} (\bm{x})}{\partial z_j} \frac{\partial \mathcal{F} (\bm{x})}{\partial z_{j'}} \right) \Sigma_{jj'} \right) .
\end{equation}


More precisely, when performing a $\tfrac{1}{S}$-expansion of all cumulants of $\mathcal{F}(\bm{X})$ and retaining only terms up to order $\tfrac{1}{S}$, we find that the expectation $\kappa_1(\mathcal{F}(\bm{X})) = \mathbb{E}[\mathcal{F}(\bm{X})]$ and the covariance $\kappa_2(\mathcal{F}(\bm{X})) = \mathrm{Cov}[\mathcal{F}(\bm{X})]$ both have non-vanishing contributions at order $\tfrac{1}{S}$ (see Appendix~\ref{appsec:cumulant-qsn-2}). In contrast, for all $M \geq 3$ the higher-order cumulants $\kappa_M(\mathcal{F}(\bm{X}))$ vanish faster, namely $\kappa_M(\mathcal{F}(\bm{X})) = o\!\left(\tfrac{1}{S}\right)$ (see Appendix~\ref{appsec:cumulant-qsn-4}).

\subsection{$(1 / S)$-expansion of function of noisy features}
\label{appsec:cumulant-qsn-1}

Before analyzing the cumulants of $\mathcal{F} (\bm{X})$, we first present an important result which will be repeatedly used in later computations. That is, for any function $\mathcal{G} (\bm{z})$, we show that $\mathcal{G} (\bm{X}) =\mathcal{G} (\bm{x} + \bm{\zeta})$ has expectation with $\frac{1}{S}$-expansion:

\begin{equation}
    \mathbb{E} [\mathcal{G} (\bm{X})] =\mathcal{G} (\bm{x}) + \frac{1}{2 S} \sum_{j, j'} \frac{\partial^2 \mathcal{G} (\bm{x})}{\partial z_j \partial z_{j'}} \bm{\Sigma}_{jj'} + o \left( \frac{1}{S} \right) . \label{apxeq:EGX}
\end{equation}

Such expression can be straightforwardly proven by writing down the Taylor expansion of $\mathcal{G} (\bm{X})$ 
\begin{align}
    \mathbb{E} [\mathcal{G} (\bm{X})] & = \mathbb{E} [\mathcal{G} (x_1 + \zeta_1, \cdots, x_K + \zeta_K)] \nonumber\\
    & = \mathbb{E} [\mathcal{G} (\bm{x})] + \sum_{m = 1}^{\infty} \sum_{j_1, \cdots, j_m} \frac{1}{m!} \mathbb{E} [\partial_{j_1} \cdots \partial_{j_m} \mathcal{G} (\bm{x}) \zeta_{j_1} \cdots \zeta_{j_m}] \nonumber\\
    & = \mathcal{G} (\bm{x}) + \sum_{m = 1}^{\infty} \sum_{j_1, \cdots, j_m} \frac{1}{m!} \partial_{j_1} \cdots \partial_{j_m} \mathcal{G} (\bm{x}) \mathbb{E} [\zeta_{j_1} \cdots \zeta_{j_m}], 
\end{align}
where $\mathbb{E} [\zeta_{j_1} \cdots \zeta_{j_m}]$ are the central moments of $\bm{X}$, which can be expressed in terms of cumulants of $\bm{\zeta}$:
\begin{align}
    & \mathbb{E} [\mathcal{G} (\bm{X})] \nonumber\\
    =~& \mathcal{G} (\bm{x}) + \sum_{m = 1}^{\infty} \sum_{j_1, \cdots, j_m} \frac{1}{m!} \partial_{j_1} \cdots \partial_{j_m} \mathcal{G} (\bm{x}) \sum_{\pi \text{ of } [m] \text{ with no single set}} \prod_{B \in \pi, | B | \geq 2} \kappa_{| B |} (\zeta_{j_i} : i \in B) \nonumber\\
    =~& \mathcal{G} (\bm{x}) + \sum_{m = 1}^{\infty} \sum_{j_1, \cdots, j_m} \frac{1}{m!} \partial_{j_1} \cdots \partial_{j_m} \mathcal{G} (\bm{x}) \sum_{\pi \text{ of } [m] \text{ with no single set}} \left( \frac{1}{S} \right)^{m - | \pi |} \prod_{B \in \pi, | B | \geq 2} \kappa_{j_{B_1} \cdots j_{B_{| B |}}} . 
\end{align}
$\kappa_{| B |} (\zeta_{j_i} : i \in B)$ represents the components $| B |$-th order tensor $\bm{\kappa}^{(| B |)}/S^{|B|-1}$, whose indices are $\{ j_i \}_{i \in B}$. We also emphasize here that $m - | \pi | \geq \left\lceil \frac{m}{2} \right\rceil$, since all $| B | = 1$ contributes nothing (first order cumulant always vanishes $\kappa_1 (\zeta_j) \equiv 0$) in summation and thus they are excluded. Then by taking the truncation at $m = 2$, we can obtain Eq.\,(\ref{apxeq:EGX}) directly.

\subsection{First and second order cumulant}
\label{appsec:cumulant-qsn-2}

Now we can perform the cumulant analysis of $\mathcal{F} (\bm{X})$. For the first order cumulant (expectation), we just simply set $\mathcal{G}=\mathcal{F}$ in Eq.\,(\ref{apxeq:EGX}):
\begin{equation}
    \mathbb{E} [\mathcal{F} (\bm{X})] =\mathcal{F} (\bm{x}) + \frac{1}{2 S} \sum_{j, j'} \frac{\partial^2 \mathcal{F} (\bm{x})}{\partial z_j \partial z_{j'}} \bm{\Sigma}_{jj'} + o \left( \frac{1}{S} \right) .
\end{equation}
Similarly, the second order moment of $\mathcal{F} (\bm{X})$ is computed by taking $\mathcal{G}=\mathcal{F}^2$ in Eq.\,(\ref{apxeq:EGX})
\begin{equation}
    \mathbb{E}_{\bm{\zeta}} [\mathcal{F} (\bm{X})^2] =\mathcal{F} (\bm{x})^2 + \frac{1}{2 S} \sum_{j, j'}^K \frac{\partial^2 (\mathcal{F} (\bm{x})^2)}{\partial z_j \partial z_{j'}} \bm{\Sigma}_{jj'} + o \left( \frac{1}{S} \right) .
\end{equation}
We need to expend the term $\frac{1}{2} \frac{\partial^2 (\mathcal{F} (\bm{x})^2)}{\partial z_j \partial z_{j'}}$:
\begin{equation}
    \frac{1}{2} \frac{\partial^2 (\mathcal{F} (\bm{x})^2)}{\partial z_j \partial z_{j'}} = \frac{\partial}{\partial z_{j'}} \left( \mathcal{F} (\bm{x}) \frac{\partial \mathcal{F} (\bm{x})}{\partial z_j} \right) = \frac{\partial \mathcal{F} (\bm{x})}{\partial z_j} \frac{\partial \mathcal{F} (\bm{x})}{\partial z_{j'}} +\mathcal{F} (\bm{x}) \frac{\partial^2 \mathcal{F} (\bm{x})}{\partial z_j \partial z_{j'}} .
\end{equation}
Hence, the covariance of $\mathcal{F} (\bm{X})$ is
\begin{align}
    & \mathrm{Cov} [\mathcal{F} (\bm{X})] =\mathbb{E} [\mathcal{F} (\bm{X})^2] -\mathbb{E} [\mathcal{F} (\bm{X})]^2 \nonumber\\
    =~& \frac{1}{S} \sum_{j, j'} \left( \frac{\partial \mathcal{F} (\bm{x})}{\partial z_j} \frac{\partial \mathcal{F} (\bm{x})}{\partial z_{j'}} +\mathcal{F} \frac{\partial^2 \mathcal{F} (\bm{x})}{\partial z_j \partial z_{j'}} \right) \bm{\Sigma}_{jj'} - \frac{1}{S} \mathcal{F} (\bm{x}) \sum_{j, j'} \frac{\partial^2 \mathcal{F} (\bm{x})}{\partial z_k \partial z_{j'}} \bm{\Sigma}_{j j'} + o \left( \frac{1}{S} \right) \nonumber\\
    =~& \frac{1}{S} \sum_{j, j'} \frac{\partial \mathcal{F} (\bm{x})}{\partial z_j} \frac{\partial \mathcal{F} (\bm{x})}{\partial z_{j'}} \bm{\Sigma}_{jj'} (\theta) + o \left( \frac{1}{S} \right) . 
\end{align}

\subsection{Higher order cumulants}
\label{appsec:cumulant-qsn-4}

The $M$-th order cumulants can be written as
\begin{equation}
    \kappa_M (\mathcal{F} (\bm{X})) = \sum_{\pi = \{ (m_l, n_l) \}} (| \pi | - 1) ! (- 1)^{| \pi | - 1} \prod_l \mathbb{E} [\mathcal{F} (\bm{X})^{m_l}]^{n_l}, \label{apxeq:kMFX}
\end{equation}
where $\pi = \{ (m_l, n_l) \}$ is a partition of $[M] := \{ 1, 2, \cdots, M \}$ where there are $n_j$ parts having $m_j$ elements, thus $| \pi | = \sum_l n_l$.

First of all, by taking $\mathcal{G}=\mathcal{F}^m$ in Eq.\,(\ref{apxeq:EGX}) we have
\begin{equation}
    \mathbb{E} [\mathcal{F} (\bm{X})^m] =\mathcal{F} (\bm{x})^m + \frac{1}{2 S} \sum_{j, j' = 1}^K \frac{\partial^2 (\mathcal{F} (\bm{x})^m)}{\partial z_j \partial z_{j'}} \bm{\Sigma}_{j j'} + o \left( \frac{1}{S} \right),
\end{equation}
where
\begin{equation}
    \frac{\partial^2 (\mathcal{F} (\bm{x})^m)}{\partial z_j \partial z_{j'}} = m (m - 1) \mathcal{F} (\bm{x})^{m - 2} \frac{\partial \mathcal{F} (\bm{x})}{\partial z_j} \frac{\partial \mathcal{F} (\bm{x})}{\partial z_{j'}} + m\mathcal{F} (\bm{x})^{m - 1} \frac{\partial^2 \mathcal{F} (\bm{x})}{\partial z_j \partial z_{j'}} .
\end{equation}
Hence, 
\begin{align}
    \mathbb{E} [\mathcal{F} (\bm{X})^m]^n =~& \mathcal{F} (\bm{x})^{m n} + \frac{m n}{2} \mathcal{F} (\bm{x})^{m n - 2} \sum_{j, j'} \left( (m - 1) \frac{\partial \mathcal{F} (\bm{x})}{\partial z_j} \frac{\partial \mathcal{F} (\bm{x})}{\partial z_{j'}} \right. \nonumber\\
    & + \left.\mathcal{F} (\bm{x}) \frac{\partial^2 \mathcal{F} (\bm{x})}{\partial z_j \partial z_{j'}} \right) \frac{\bm{\Sigma}_{j j'}}{S} + o \left( \frac{1}{S} \right),
\end{align}
By dropping $o \left( \frac{1}{S} \right)$, we get (we use the fact $\sum_l m_l n_l = M$)
\begin{align}
    & \prod_j \mathbb{E} [\mathcal{F} (\bm{X})^{m_l}]^{n_l} \nonumber\\
    =~& \mathcal{F} (\bm{x})^{\sum_l m_l n_l} + \frac{\mathcal{F} (\bm{x})^{\sum_l m_l n_l - 2}}{2} \sum_{j, j'} \left( \left( \sum_l m_l (m_l - 1) n_l \right) \frac{\partial \mathcal{F} (\bm{x})}{\partial z_j} \frac{\partial \mathcal{F} (\bm{x})}{\partial z_{j'}} \right. \nonumber\\
    & \left. + \left( \sum_l m_l n_l \right) \mathcal{F} (\bm{x}) \frac{\partial^2 \mathcal{F} (\bm{x})}{\partial z_j \partial z_{j'}} \right) \frac{\bm{\Sigma}_{j j'}}{S} \nonumber\\
    =~& \mathcal{F} (\bm{x})^M + \frac{\mathcal{F} (\bm{x})^{M - 2}}{2} \sum_{j, j'} \left( \left( \sum_l m_l (m_l - 1) n_l \right) \frac{\partial \mathcal{F} (\bm{x})}{\partial z_j} \frac{\partial \mathcal{F} (\bm{x})}{\partial z_{j'}} \right. \nonumber\\
    & \left.+ \left( \sum_l m_l n_l \right) \mathcal{F} (\bm{x}) \frac{\partial^2 \mathcal{F} (\bm{x})}{\partial z_j \partial z_{j'}} \right) \frac{\bm{\Sigma}_{j j'}}{S} . 
\end{align}
Now we use the identities
\begin{align}
    \sum_{\pi = \{ (m_l, n_l) \}} (| \pi | - 1) ! (- 1)^{| \pi | - 1} & = 0, \\
    \sum_{\pi = \{ (m_l, n_l) \}} (| \pi | - 1) ! (- 1)^{| \pi | - 1} \sum_l m_l n_l & = \delta_{1 M}, \\
    \sum_{\pi = \{ (m_l, n_l) \}} (| \pi | - 1) ! (- 1)^{| \pi | - 1} \sum_l \frac{m_l (m_l - 1) n_l}{2} & = \delta_{2 M}, 
\end{align}
we obtain that all term in Eq.\,(\ref{apxeq:kMFX}) vanishes other than $o \left( \frac{1}{S} \right)$, and therefore $\kappa_M (\mathcal{F} (\bm{X})) = o \left( \frac{1}{S} \right)$ for all $M \geq 3$.


\section{Post-reservoir Loss Function}
\label{appsec:post-rc-loss}

In this section, we provide a general formalism to analyze the post-reservoir loss function, and explicitly show that the post-reservoir loss in Eq.\,(\ref{eq:post_reservoir_loss}) is approximately Gaussian, that is $\hat{L}(\bm{\theta}) \sim \mathcal{N}(\mu(\bm{\theta}), s(\bm{\theta}))$ where $\mu(\bm{\theta})$ and $s(\bm{\theta})$ to be determined.

To simplify the notation in the derivation, we define a loss function $\mathcal{L}(\bm{w}, \bm{z})$, where $\bm{z}$ generically denotes the feature vector. When quantum shot noise is included, we use $\bm{z} \leftarrow \bm{X}(\bm{u})$, and the loss is given by $\mathcal{L}(\bm{w}, \bm{X}(\bm{u}))$.
There is a unique set of weights $\bm{w}^{\ast}$ s.t.
\begin{equation}
    \bm{w}^{\ast} (\bm{\theta}) = \mathrm{argmin}_{\bm{w}}  \frac{1}{N} \sum_n \mathcal{L} (\bm{w}, \bm{X} (\bm{u}^{(n)})) .
\end{equation}
This is equivalently solved by equations
\begin{equation}
    \sum_{n'} \frac{\partial \mathcal{L}}{\partial w_j} (\bm{w}^{\ast}, \bm{z}^{(n')}) = 0, 
    \label{apxeq:dLdw}
\end{equation}
by substituting $(\bm{z}^{(1)}, \cdots, \bm{z}^{(N)}) \leftarrow (\bm{X} (\bm{u}^{(1)}), \cdots, \bm{X} (\bm{u}^{(N)}))$. It gives a natural definition of an $(N K)$-variate function $\bm{w}^{\ast} (\bm{z}^{(1)}, \cdots, \bm{z}^{(N)})$ and
\begin{equation}
    \mathcal{L}^{\ast} (\bm{z}^{(1)}, \cdots, \bm{z}^{(N)}) = \frac{1}{N} \sum_{n' = 1}^N \mathcal{L} (\bm{w}^{\ast} (\bm{z}^{(1)}, \cdots, \bm{z}^{(N)}), \bm{z}^{(n')}) .
\end{equation}
We notice that $\mathcal{L}^{\ast} (\bm{X} (\bm{u}^{(1)}), \cdots, \bm{X} (\bm{u}^{(N)})) = \hat{L} (\bm{\theta})$ as defined in Eq.\,(\ref{eq:post_reservoir_loss}). Hence in this section, we let function $\mathcal{G}=\mathcal{L}^{\ast}$ in Eq.\,(\ref{apxeq:EGX}). The underlying random variable is $(N K)$-variate $(\bm{X} (\bm{u}^{(1)}), \cdots, \bm{X} (\bm{u}^{(N)}))$, whose second order cumulant is block-wise diagonal:
\begin{equation}
    \bm{\kappa}^{(2)} = \mathrm{diag} \{ \bm{\Sigma} (\bm{u}^{(n)}) \} 
    = \left(\begin{array}{ccc}
        \bm{\Sigma} (\bm{u}^{(1)}) & \cdots & \bm{0}\\
        \vdots & \ddots & \vdots\\
        \bm{0} & \cdots & \bm{\Sigma} (\bm{u}^{(N)})
    \end{array}\right),
\end{equation}
because $\bm{X} (\bm{u}^{(n)})$ and $\bm{X} (\bm{u}^{(n')})$ are independent for any $n \neq n'$. Then by applying Eq.\,(\ref{apxeq:EGX}), we obtain that $\mathcal{L}^{\ast} (\bm{X} (\bm{u}^{(1)}), \cdots, \bm{X} (\bm{u}^{(N)})) = \hat{L} (\bm{\theta})$ can be approximated by Gaussian
\begin{equation}
    \mathcal{N} \left( \mathcal{L}^{\ast} (\{ \bm{x} (\bm{u}^{(n)}) \}) + \frac{1}{2 S} \sum_{n, j, j'} \frac{\partial^2  \mathcal{L}^{\ast}}{\partial z^{(n)}_j \partial z^{(n)}_{j'}}  \bm{\Sigma}_{j j'} (\bm{u}^{(n)}), \frac{1}{S} \sum_{n, j, j'} \left( \frac{\partial \mathcal{L}^{\ast}}{\partial z^{(n)}_j} \frac{\partial \mathcal{L}^{\ast}}{\partial z^{(n)}_{j'}} \right) \bm{\Sigma}_{j j'} (\bm{u}^{(n)}) \right) . \label{apxeq:GaussLast}
\end{equation}

Although $\mathcal{L}^{\ast} (\bm{z}^{(1)}, \cdots, \bm{z}^{(N)})$ is a highly complex function, within the RGD framework, both $\frac{\partial \mathcal{L}^{\ast}}{\partial z^{(n)}_j}$ and $\frac{\partial^2 \mathcal{L}^{\ast}}{\partial z^{(n)}_j \partial z^{(n)}_{j'}}$ can be drastically simplified. This follows from the fact that $\bm{w}^{\ast}(\bm{\theta})$ minimizes the loss function for fixed internal parameters, $\bm{\theta}$, and satisfies Eq.~(\ref{apxeq:dLdw})

\subsection{First-order derivative of $\mathcal{L}^{\ast}$}

Since $\sum_{n'} \frac{\partial \mathcal{L}}{\partial w_k} (\bm{w}^{\ast}, \bm{z}^{(n')}) = 0$ for all $k$, we have
\begin{align}
    \frac{\partial \mathcal{L}^{\ast}}{\partial z^{(n)}_j} & = \frac{1}{N} \left( \sum_{n', k} \frac{\partial \mathcal{L}}{\partial w_k} (\bm{w}^{\ast}, \bm{z}^{(n')}) \frac{\partial w^{\ast}_k}{\partial z^{(n)}_j} (\bm{z}^{(1)}, \cdots, \bm{z}^{(N)}) + \frac{\partial \mathcal{L}}{\partial z_j} (\bm{w}^{\ast}, \bm{z}^{(n)}) \right) \nonumber\\
    & = \frac{1}{N} \left( \sum_k \underbrace{\left( \sum_{n'} \frac{\partial \mathcal{L}}{\partial w_k} (\bm{w}^{\ast}, \bm{z}^{(n')}) \right)}_{= 0} \frac{\partial w^{\ast}_k}{\partial z^{(n)}_j} + \frac{\partial \mathcal{L}}{\partial z_j} (\bm{w}^{\ast}, \bm{z}^{(n)}) \right) \nonumber\\
    & = \frac{1}{N} \frac{\partial \mathcal{L}}{\partial z_j} (\bm{w}^{\ast}, \bm{z}^{(n)}) . \label{apxeq:1stDLast}
\end{align}

\subsection{Second-order derivative of $\mathcal{L}^{\ast}$}

By taking derivative of Eq.\,(\ref{apxeq:1stDLast}) with respect to $z_{j'}$. We notice that
\begin{equation}
    \frac{\partial^2 \mathcal{L}^{\ast}}{\partial z^{(n)}_j \partial z^{(n)}_{j'}} = \frac{1}{N} \left( \sum_l \frac{\partial^2 \mathcal{L}}{\partial z_j \partial w_l} (\bm{w}^{\ast}, \bm{z}^{(n)}) \frac{\partial w^{\ast}_l}{\partial z^{(n)}_{j'}} + \frac{\partial^2 \mathcal{L}}{\partial z_j \partial z_{j'}} (\bm{w}^{\ast}, \bm{z}^{(n)}) \right) .
\end{equation}
The only unknown quantity $\frac{\partial w^{\ast}_l}{\partial z^{(n)}_{j'}}$ is the changing rate of optimal weights with respect to $z^{(n)}_{j'}$. It can be computed by taking derivative of Eq.\,(\ref{apxeq:dLdw}) with respect to $z_j$. That is, $\frac{\partial}{\partial z^{(n)}_j} \left( \sum_{n'} \frac{\partial \mathcal{L}}{\partial w_k} (\bm{w}^{\ast}, \bm{z}^{(n')}) \right) = 0$ implies
\begin{equation}
    \sum_{n', l} \frac{\partial \mathcal{L}}{\partial w_l \partial w_k} (\bm{w}^{\ast}, \bm{z}^{(n')}) \frac{\partial w^{\ast}_l}{\partial z^{(n)}_j} + \frac{\partial \mathcal{L}}{\partial z_j \partial w_k} (\bm{w}^{\ast}, \bm{z}^{(n)}) = 0 \quad \Rightarrow \quad \sum_{j'} \mathbf{H}_{k l} \frac{\partial w^{\ast}_{l}}{\partial z^{(n)}_j} = -\mathbf{K}^{(n)}_{k j},
\end{equation}

where we define $\mathbf{H}_{k l} = \sum_{n'} \frac{\partial^2 \mathcal{L}}{\partial w_k \partial w_l} (\bm{w}^{\ast}, \bm{z}^{(n')})$ and $\mathbf{K}^{(n)}_{k j} = \frac{\partial^2 \mathcal{L}}{\partial w_k \partial z_j} (\bm{w}^{\ast}, \bm{z}^{(n)})$. It is equivalently $\frac{\partial w^{\ast}_l}{\partial z^{(n)}_{j}} = - (\mathbf{H}^{- 1} \mathbf{K}^{(n)})_{l, j}$ and thus
\begin{equation}
    \frac{\partial^2 \mathcal{L}^{\ast}}{\partial z^{(n)}_j \partial z^{(n)}_{j'}} = \frac{1}{N} \left( \frac{\partial^2 \mathcal{L}}{\partial z_j \partial z_{j'}} (\bm{w}^{\ast}, \bm{z}^{(n)}) - (\mathbf{K}^{(n) T} \mathbf{H}^{- 1} \mathbf{K}^{(n)})_{j, j'} \right) . \label{apxeq:2ndDLast}
\end{equation}
Finally, we plug Eq.\,(\ref{apxeq:1stDLast}) and Eq.\,(\ref{apxeq:2ndDLast}) into Eq.\,(\ref{apxeq:GaussLast}), and achieves the result the Eq.\,(\ref{apxeq:GaussLast}).


\section{Reservoir gradient-free optimization}
\label{sec:reservoir_direct}
As discussed in the main text, jointly optimizing the full sensing pipeline including both the internal parameters of the quantum circuit and the final estimator, presents significant computational challenges due to the high-dimensional, non-convex, and often noisy nature of the optimization landscape. In this section we provide more details on the post-reservoir gradient-free optimization strategy we developed and used to identify the optimal circuit parameters and estimators for the sensing tasks presented in the main text. 

Inspired by the DIviding RECTangle (DIRECT) method \cite{Jones1993,article_direkt,nicholas2015dividing,article_direkt2} our algorithm strategically selects the evaluation points of the post-reservoir loss function defined as:
\begin{equation}
    \hat{L}(\bm{\theta}) = \min_{\bm{w} \in \mathbb{R}^K} \, \hat{l}(\bm{\theta}, \bm{w}), 
\end{equation}
which corresponds to Eq.~\eqref{eq:empir_loss} of the main text. Here, 
\begin{align}
\hat{l}(\bm{\theta}, \bm{w}) &= \frac{1}{N} \sum_{n=1}^N \mathcal{L}(\bm{w}, \bm{X}(\bm{u}^{(n)})),
\end{align}
is the empirical loss computed with finite samples, $N$ and shots, $S$. Crucially, for any fixed choice of system parameters $\bm{\theta}$, the optimization of the linear estimator weights, $\bm{w}$, is convex allowing for efficient computation. To guide the selection of the evaluation points, the algorithm maintains an internal representation of the objective loss function in the form of a surrogate model. Specifically, we employ a Gaussian process (GP)~\cite{p947-r9l1} metamodel which provides both a predictive mean and an uncertainty estimate over the objective function. Such probabilistic framework enables informed exploration of the parameter space by balancing exploitation of low-loss regions and exploration of uncertain areas, thereby improving sample efficiency during optimization. At each optimization step, the next evaluation point is selected based on a scoring criterion that combines both exploration and exploitation. This score is constructed using the predictive mean and variance of the GP metamodel, with tunable weights that control the relative importance of these two objectives. Importantly, these weights can be adapted dynamically across epochs, allowing the algorithm to emphasize exploration in the early stages and gradually shift toward exploitation as convergence is approached.

For the successful implementation of a metamodel-based optimization approach, it is essential to select a surrogate ansatz that offers sufficient flexibility while capturing the essential features of the loss landscape. The surrogate must be capable of modeling relevant variations in the objective function across the parameter space without overfitting to noise or sparse observations. In our case, the hyperparameters of the Gaussian process metamodel are determined via maximum-likelihood estimation, ensuring that the model remains well-calibrated to the observed data throughout the optimization. 

To gain further insight into the structure of the target loss function, we begin by analyzing the behavior of a generic observable $\hat{\mathcal{O}}$ as a function of a system parameter $\theta$:
\begin{align}
    g(\theta) = \bra{\psi_0} \mathcal{U}^\dagger(\theta) \hat{\mathcal{O}} \mathcal{U}(\theta) \ket{\psi_0}.
\end{align}
Here, $\mathcal{U}(\theta) = e^{-i \theta \hat{G}}$ denotes a unitary operator generated by the Hermitian operator $\hat{G}$, and $\ket{\psi_0}$ is the initial state while without loss of generality, we focus on the dependence on a single parameter $\theta$.
Since $\hat{G}$ is Hermitian, it can be diagonalized as $\hat{G} = \sum_j \lambda_j \ket{j}\bra{j}$, where $\{\lambda_j\}$ are real eigenvalues and $\{\ket{j}\}$ form an orthonormal eigenbasis. This leads to the decomposition
\begin{align}
    \mathcal{U}(\theta) = \sum_j e^{-i \theta \lambda_j} \ket{j}\bra{j},
\end{align}
and consequently,
\begin{align}
    g(\theta) &= \bra{\psi_0} \left( \sum_k e^{i \theta \lambda_k} \ket{k}\bra{k} \right) \hat{\mathcal{O}} \left( \sum_l e^{-i \theta \lambda_l} \ket{l} \bra{l} \right) \ket{\psi_0} \\
    &= \sum_{k,l} c_k^* c_l \bra{k} \hat{\mathcal{O}} \ket{l} \, e^{i \theta (\lambda_k - \lambda_l)},
\end{align}
where $c_l = \braket{l}{\psi_0}$ are the expansion coefficients of the initial state in the eigenbasis of $\hat{G}$.
As a result, $g(\theta)$ can be expressed as a finite Fourier series:
\begin{align}
    g(\theta) = \sum_{m > 0} A_m \cos(m \theta) + B_m \sin(m \theta),
\end{align}
where the frequencies $m$ correspond to the distinct values of the eigenvalue differences $\lambda_k - \lambda_l$ of the generator $\hat{G}$, and the coefficients $A_m$, $B_m$ depend on the matrix elements of $\hat{\mathcal{O}}$ and the initial state $\ket{\psi_0}$.

If $\hat{G}$ has only two distinct eigenvalues, $+r$, $-r$, then 
\begin{align}
    g(\theta)=A\cos(2r \theta)+B \sin(2r \theta)
\end{align}
and 
\begin{align}
    \frac{dg(\theta)}{d \theta}=-2rA \sin(2 r \theta)+2rB \cos(2 r \theta).
\end{align}
With some simple algebra it is straightforward to obtain that 
\begin{align}
    \frac{dg(\theta)}{d \theta}= \frac{2r}{2 \sin(2s )} \left[g(\theta+\frac{s}{r}) - g(\theta-\frac{s}{r}) \right].
\end{align}
If we choose $s=\pi/4$ then we obtain the parameter- shift rule~\cite{PhysRevA.98.032309,PhysRevA.99.032331}
\begin{align}
    \frac{dg(\theta)}{d \theta}= r \left[g(\theta+\frac{\pi}{4r}) - g(\theta-\frac{\pi}{4r}) \right].
\end{align}
Finally, this analysis can be extended to cases where the observable is obtained after the application of a sequence of unitary operations, each depending on a different parameter. Specifically, we consider
\begin{align}
    g(\bm{\theta}) = \bra{\psi_0} \left( \prod_k \mathcal{U}_k^{\dagger}(\theta_k) \right) \hat{\mathcal{O}} \left( \prod_k \mathcal{U}_k(\theta_k) \right) \ket{\psi_0},
\end{align}
which corresponds to the structure of the sensing task discussed in the main text. In such cases, the resulting function $g(\bm{\theta})$ can be expressed as a product of Fourier series in each parameter:
\begin{align}
    g(\bm{\theta}) = \prod_i \sum_{m_i} \left[ A_{m_i} \cos(m_i \theta_i) + B_{m_i} \sin(m_i \theta_i) \right],
\end{align}
where $m_i$ denotes the possible differences between eigenvalues of the generator associated with parameter $\theta_i$.

\begin{figure}[h]
    \centering
    \includegraphics[scale=1.0]
    {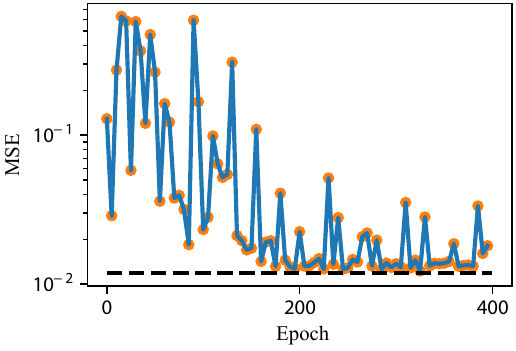}
    \caption{Optimization trajectory. Bayesian mean squared error as the number of iteration for finding the optimal circuit and estimator for $n_{\rm en}=1, n_{\rm de}=2$, $S=1$, and $L=32$ (see main text). The optimization is initialized from a preliminary estimate obtained via Gauss--Hermite quadrature with $n=75$ points, corresponding to boxes spanning $50\%$ to $150\%$ of the initial optimum leading to a fast convergence. The algorithm is guaranteed to find the optimal solution (shown in black dashed line) since not only exploits promising values of the parameter space but keeps exploring new parts that are less explored. The occasional jumps to higher BMSE values arise from mis-approximations of the true loss function by the surrogate model, which are subsequently corrected as the optimization progresses.}
    \label{fig:error_tr_epoch}
\end{figure}

This functional form highlights that $g(\bm{\theta})$ resides within a low-bandwidth, structured function space. Therefore, selecting a surrogate model ansatz that reflects this Fourier structure can significantly enhance the efficiency and accuracy of the optimization process. In Fig.~\ref{fig:error_tr_epoch} we demonstrate the numerical convergence of our algorithm for estimating the unknown parameter $u$ and for a gaussian prior distribution as a function of the number of iterations (epochs). The optimization is initialized within a search region restricted to boxes spanning from $50\%$ to $150\%$ of a preliminary optimal value. This initial estimate is obtained from an extensive exploration of the loss landscape using Gauss-Hermite quadrature integration, which is applicable in the case of Gaussian prior distributions of the unknown parameter $u$.

Within this approach, any Gaussian integral of the form $\int_{-\infty}^{\infty} dx e^{-x^2} f(x)$ can be approximated by the Gauss-Hermite quadrature rule

\begin{align}
\int_{-\infty}^{\infty} dx e^{-x^2} f(x) \approx \sum_{i=1}^n \beta_i f(x_i),
\end{align}
where $n$ denotes the number of quadrature points $x_i$ are the roots of the Hermite polynomial $H_n(x)$ and $\beta_i$ are the corresponding weights. In the results presented in this work, we fix $n=75$, which already provides sufficient accuracy while drastically reducing the number of evaluations of the loss function, thereby making the initial search highly efficient.

\section{Noise analysis}
\label{app:noise_analysis}

In this section we analyze in more detail the different noise sources considered in the main text. In particular we focus on i) readout errors and (ii) gate imperfections that drive the dynamics outside the symmetric subspace and we demonstrate that our end-to-end optimization framework remain robust in the presence of such imperfections.

We focus on these two error sources because, in contrast to decoherence, they can be partially mitigated through the co-optimization of the parametrized quantum circuit and estimator. In particular, the end-to-end framework can adapt the entangling, decoding, and estimation stages in order to identify noise-resilient feature combinations and measurement strategies.

\subsection{Microscopic model of readout errors}
\label{app:readout_errors}

Here, we provide a microscopic motivation for the effective readout-error model used in the main text. Since the measurement considered in the collective-spin setting is the total magnetization, readout imperfections can be incorporated as a classical noisy channel acting on the ideal probability distribution of magnetization outcomes. If $P_m(u)$ denotes the ideal probability of obtaining magnetization outcome $m$ for an input parameter $u$, the experimentally observed probability distribution is written as
\begin{equation}
\widetilde{P}_{m}(u)=\sum_{m'} \Gamma_{m,m'} P_{m'}(u),
\label{eq:noisy_readout_channel_app}
\end{equation}
where $\Gamma_{m,m'}$ is the conditional probability that the ideal outcome $m'$ is recorded as $m$. In the main text, we model this channel through the effective Gaussian kernel
\begin{equation}
\Gamma_{m,m'}=\frac{\exp\!\left[-(m-m')^2/(2\sigma_{\rm ro}^2)\right]}{\sum_{m''}
\exp\!\left[-(m''-m')^2/(2\sigma_{\rm ro}^2)\right]
},
\label{eq:gaussian_readout_kernel_app}
\end{equation}
where $\sigma_{\rm ro}$ controls the readout resolution. This model captures the finite resolution of collective-spin measurements, for example due to imperfect fluorescence collection, thresholding errors, or coarse-grained magnetization detection. We now show how such a kernel naturally emerges from independent single-qubit assignment errors after coarse graining to the total magnetization.

Consider a system of $L$ qubits. Let $x_i\in\{0,1\}$ denote the true state of qubit $i$ and $y_i\in\{0,1\}$ the corresponding recorded measurement outcome. We assume independent readout errors on each qubit, with
\begin{align}
P(y_i=0|x_i=1) &= p_1, \\
P(y_i=1|x_i=0) &= p_0,
\end{align}
where $p_1$ is the probability of incorrectly assigning an excited qubit as being in the ground state, while $p_0$ is the probability of incorrectly assigning a ground-state qubit as being excited. The corresponding correct-assignment probabilities are $1-p_1$ and $1-p_0$.

Suppose that the true bitstring contains $n_\uparrow$ spins in the state $1$. The measured number of spins in the state $1$, denoted by $\widetilde{n}_\uparrow$, receives contributions from two independent stochastic processes: correctly identified excited spins and incorrectly identified ground-state spins. If $u$ of the $n_\uparrow$ true excited spins are correctly identified, and $k-u$ of the $L-n_\uparrow$ true ground-state spins are incorrectly assigned as excited, then the probability of measuring $\widetilde{n}_\uparrow=k$ is
\begin{equation}
P(\widetilde{n}_\uparrow=k|n_\uparrow)
=
\sum_u
\binom{n_\uparrow}{u}
(1-p_1)^u p_1^{n_\uparrow-u}
\binom{L-n_\uparrow}{k-u}
p_0^{k-u}
(1-p_0)^{L-n_\uparrow-(k-u)} .
\label{eq:microscopic_readout_kernel_nup}
\end{equation}
The sum is restricted to values of $u$ for which all binomial coefficients are well defined, i.e.
\begin{equation}
\max\{0,k-(L-n_\uparrow)\}
\leq u \leq
\min\{k,n_\uparrow\}.
\end{equation}
Equation~\eqref{eq:microscopic_readout_kernel_nup} is the exact readout-error kernel at the level of the measured excitation number.

It is useful to rewrite this result in terms of magnetization variables,
\begin{equation}
m = n_\uparrow-\frac{L}{2},
\qquad
\widetilde{m}
=
\widetilde{n}_\uparrow-\frac{L}{2}.
\label{eq:magnetization_definition_app}
\end{equation}
From Eq.~\eqref{eq:microscopic_readout_kernel_nup}, the conditional mean of the measured magnetization is
\begin{equation}
\mathbb{E}[\widetilde{m}|m]
=
(1-p_0-p_1)m
+
\frac{L}{2}(p_0-p_1),
\label{eq:readout_mean_general_app}
\end{equation}
while the conditional variance is
\begin{equation}
\mathrm{Var}[\widetilde{m}|m]
=
\left(m+\frac{L}{2}\right)p_1(1-p_1)
+
\left(\frac{L}{2}-m\right)p_0(1-p_0).
\label{eq:readout_variance_general_app}
\end{equation}
Thus, asymmetric readout errors have two effects. First, they reduce the contrast of the measured magnetization through the prefactor $1-p_0-p_1$. Second, they introduce a bias proportional to $p_0-p_1$, which shifts the measured magnetization even when the true magnetization is zero.

For the symmetric assignment-error model, $p_0=p_1=p$, the bias term vanishes and the expressions simplify to
\begin{align}
\mathbb{E}[\widetilde{m}|m]
&=
(1-2p)m,
\label{eq:readout_mean_symmetric_app}
\\
\mathrm{Var}[\widetilde{m}|m]
&=
L p(1-p).
\label{eq:readout_variance_symmetric_app}
\end{align}
Therefore, symmetric readout errors do not shift the origin of the magnetization distribution, but they attenuate the signal by the factor $(1-2p)$ and broaden the measured distribution with a variance proportional to $L p(1-p)$. In the experimentally relevant regime $p\ll 1$, the contrast reduction is small, but it is nevertheless a systematic effect distinct from ordinary finite-shot sampling noise.

For sufficiently large $L$, and $p\ll 1$ as it is typically the case for neutral-atoms and trapped-ions setups the conditional distribution of the measured magnetization approaches a Gaussian by the central-limit theorem as described in q.~\eqref{eq:gaussian_readout_kernel_app} with $\sigma_{\rm ro}^2 \sim Lp(1-p)$. In the symmetric case this gives the approximate coarse-grained kernel

For completeness, we also state the corresponding bitstring-level form of the readout channel. Let $\mathbf{x}=(x_1,\dots,x_L)$ and $\mathbf{y}=(y_1,\dots,y_L)$ denote the true and measured bitstrings. Under independent assignment errors, the conditional probability factorizes as
\begin{equation}
\Gamma_{\mathbf{y},\mathbf{x}}
=
\prod_{i=1}^{L} P(y_i|x_i).
\label{eq:readout_errors_bitstring_app}
\end{equation}
For symmetric readout errors with probability $p$, this becomes
\begin{equation}
\Gamma_{\mathbf{y},\mathbf{x}}
=
(1-p)^{L-d_H(\mathbf{x},\mathbf{y})}
p^{d_H(\mathbf{x},\mathbf{y})},
\label{eq:hamming_readout_kernel_app}
\end{equation}
where
\begin{equation}
d_H(\mathbf{x},\mathbf{y})
=
\sum_{i=1}^{L}|x_i-y_i|
\end{equation}
is the Hamming distance between the true and measured bitstrings. Thus, the probability of recording a wrong bitstring is exponentially controlled by the number of flipped measurement outcomes. When the final observable is the total magnetization, Eq.~\eqref{eq:hamming_readout_kernel_app} can be coarse grained over all bitstrings with fixed Hamming weight. This coarse graining yields the magnetization-level kernel in Eq.~\eqref{eq:microscopic_readout_kernel_nup}, which approaches the Gaussian form in Eq.~\eqref{eq:gaussian_readout_kernel_app} for large $L$.

\subsection{Gate imperfections}
\label{app:gate_errors}

We now consider coherent gate imperfections that modify the quantum dynamics itself. This is qualitatively different from the readout errors discussed above. Readout errors act after the quantum evolution and can therefore be represented as a classical noisy channel acting on the final measurement probabilities. However, gate imperfections change the state generated by the sensing protocol. In particular, imperfections that break permutation symmetry can drive the wavefunction outside the symmetric Dicke manifold.

In the ideal collective-spin model used in the main text, the entangling and decoding operations are generated by collective spin operators. Therefore, if the initial state belongs to the fully symmetric subspace, the dynamics remains confined to this subspace. This is what allows the simulation of $L$ qubits in the reduced Hilbert space of dimension $L+1$, rather than in the full Hilbert space of dimension $2^L$. Once permutation symmetry is broken, this reduction is no longer valid. For this reason, the gate-imperfection analysis is performed in the full Hilbert space. Because of the exponential scaling of the Hilbert-space dimension, we restrict this part of the analysis to a representative system size with $L=4$, $n_{\rm en}=1$, $n_{\rm de}=2$, and prior width $\sigma_u=0.7981$.

As a reference point, we first perform the optimization in the full Hilbert space without adding symmetry-breaking gate disorder. This allows us to verify that indeed the end-to-end optimization naturally selects solutions close to the symmetric Dicke manifold even when the full Hilbert space is available and the optimized dynamics remains predominantly supported in the symmetric subspace. Thus, the solution found by the optimization algorithm is not an artifact of explicitly imposing the symmetric-subspace restriction; rather, the optimization naturally recovers the collective-spin structure of the ideal model.

To analyze the structure of the optimized states, we consider their distribution over Hamming-weight sectors. For a computational basis state
\begin{equation}
\ket{\mathbf{x}}
=
\ket{x_1,\dots,x_L},
\qquad
x_i\in\{0,1\},
\end{equation}
the Hamming weight is defined as
\begin{equation}
w_H(\mathbf{x})
=
\sum_{i=1}^{L} x_i .
\label{eq:hamming_weight_app}
\end{equation}
The Hamming weight therefore counts the number of spins in the excited state. The projector onto the subspace with fixed Hamming weight $w$ is
\begin{equation}
\Pi_w
=
\sum_{\mathbf{x}:w_H(\mathbf{x})=w}
\ket{\mathbf{x}}\bra{\mathbf{x}} .
\label{eq:hamming_projector_app}
\end{equation}
For a state $\ket{\psi(u)}$, the population in the Hamming-weight sector $w$ is then
\begin{equation}
P_w(u)
=
\bra{\psi(u)}\Pi_w\ket{\psi(u)} .
\label{eq:hamming_population_app}
\end{equation}
This quantity provides a useful diagnostic of how the optimized protocol distributes population among different excitation-number sectors.

The connection to the symmetric subspace is obtained by recalling that the Dicke state with $w$ excitations is
\begin{equation}
\ket{D_w^{(L)}}
=
\binom{L}{w}^{-1/2}
\sum_{\mathbf{x}:w_H(\mathbf{x})=w}
\ket{\mathbf{x}} .
\label{eq:dicke_state_app}
\end{equation}
The fully symmetric subspace is spanned by the $L+1$ Dicke states
\begin{equation}
\left\{
\ket{D_w^{(L)}} \right\}_{w=0}^{L}.
\end{equation}
Thus, a general state in the symmetric manifold can be written as
\begin{equation}
\ket{\psi_{\rm sym}(u)}
=
\sum_{w=0}^{L}
c_w(u)\ket{D_w^{(L)}} .
\label{eq:symmetric_state_app}
\end{equation}
The corresponding projector onto the symmetric subspace is
\begin{equation}
\Pi_{\rm sym}
=
\sum_{w=0}^{L}
\ket{D_w^{(L)}}\bra{D_w^{(L)}} .
\label{eq:symmetric_projector_app}
\end{equation}
We quantify the total support on the symmetric manifold by
\begin{equation}
W_{\rm sym}(u)
=
\bra{\psi(u)}\Pi_{\rm sym}\ket{\psi(u)} ,
\label{eq:symmetric_weight_app}
\end{equation}
and the corresponding leakage outside the symmetric subspace by
\begin{equation}
W_{\rm leak}(u)
=
1-W_{\rm sym}(u).
\label{eq:symmetric_leakage_app}
\end{equation}
The Hamming-weight distribution $P_w(u)$ and the symmetric-subspace weight $W_{\rm sym}(u)$ provide complementary information. The former shows how population is distributed over excitation-number sectors, while the latter directly measures whether the state remains in the fully symmetric Dicke manifold.

We now introduce gate imperfections in the one-axis-twisting operations. In the ideal collective model, one-axis twisting around the axis $\alpha\in\{x,y,z\}$ is generated by

\begin{equation}
T_\alpha(\phi)
=
\exp\!\left[-i\phi J_\alpha^2\right],
\label{eq:ideal_twisting_app}
\end{equation}

where $J_\alpha
=\frac{1}{2} \sum_{i=1}^{L} \hat{\sigma}_{\alpha}^{(i)}$. Using this definition, we can write
\begin{equation}
J_\alpha^2
=
\frac{1}{4}
\sum_{i,j=1}^{L}
\hat{\sigma}_{\alpha}^{(i)}
\hat{\sigma}_{\alpha}^{(j)}
=
\frac{L}{4}\mathbb{I}
+
\frac{1}{2}
\sum_{i<j}
\hat{\sigma}_{\alpha}^{(i)}
\hat{\sigma}_{\alpha}^{(j)} .
\label{eq:collective_twisting_pairwise_app}
\end{equation}
Therefore, up to an irrelevant global phase, ideal one-axis twisting corresponds to an all-to-all two-body interaction with identical pairwise couplings. To model imperfect all-to-all interactions, we replace the ideal twisting operation by
\begin{equation}
\mathcal{T}_{\alpha}(\phi)
=
\exp\!\left[-i\phi H_{\alpha}^{\rm ge}\right],
\label{eq:gate_error_twist_app}
\end{equation}
with
\begin{equation}
H_{\alpha}^{\rm ge}
=
\sum_{i<j}
J_{ij}^{(\alpha)}
\hat{\sigma}_{\alpha}^{(i)}
\hat{\sigma}_{\alpha}^{(j)} .
\label{eq:gate_error_hamiltonian_app}
\end{equation}
The imperfect pairwise couplings are taken to be
\begin{equation}
J_{ij}^{(\alpha)}
=
J+\delta J_{ij}^{(\alpha)},
\label{eq:gate_error_couplings_app}
\end{equation}
where the fluctuations are sampled independently from a normal distribution,
\begin{equation}
\delta J_{ij}^{(\alpha)}
\sim
\mathcal{N}(0,\eta^2).
\label{eq:gate_error_distribution_app}
\end{equation}
Here $\eta$ controls the strength of the gate disorder. Physically, this model captures static calibration errors, imperfect control fields, and spatial inhomogeneities in the pairwise interaction strengths. In the simulations, the disorder realization is kept fixed throughout training and testing, corresponding to coherent hardware imperfections rather than stochastic gate noise that changes from shot to shot.

For $\eta=0$, all pairwise couplings are identical and the Hamiltonian in Eq.~\eqref{eq:gate_error_hamiltonian_app} reduces to the collective twisting Hamiltonian in Eq.~\eqref{eq:collective_twisting_pairwise_app}, up to an overall constant and a rescaling of the twisting angle. For $\eta>0$, the couplings depend on the pair $(i,j)$ and the Hamiltonian no longer commutes with arbitrary permutations of the qubits. The evolution therefore does not preserve the symmetric Dicke manifold and can generate leakage into nonsymmetric sectors of the full Hilbert space.

The robustness against such imperfection is shown in Fig.~\ref{fig:gate_imperfections} of the main text, demonstrating that the end-to-end procedure adapts to the actual imperfect dynamics and identifies sensing strategies that remain effective even when the ideal permutation symmetry is partially broken.

\section{Extension to arbitrary tasks}
\label{app:arbitrary_tasks}

In this Section, we analyze the expressive structure of the circuit ans\"atze considered for the spiral-classification task in Sec.~\ref{sec:arbitrary_tasks}. In particular, we show how the choice of input embedding constrains the class of input-dependent functions that can be generated by the quantum circuit and thereby provides an analytical explanation for the qualitatively different decision boundaries observed in the main text.

A useful perspective for understanding the expressive power of parametrized quantum circuits is provided by their Fourier representation. For Hamiltonian data encodings, expectation values and measurement probabilities generated by a finite-dimensional quantum circuit can generally be expressed as finite Fourier-type expansions, with the accessible frequencies determined by the spectra of the corresponding encoding generators~\cite{PhysRevA.103.032430,PhysRevA.107.062612}. The trainable part of the circuit determines how these frequencies are combined and, depending on the circuit architecture, can further constrain the Fourier coefficients that are actually accessible. Repeated or more flexible data encodings can consequently enlarge the effective function class available to a quantum model. 

As explained in the main text we start with employing an asatz similar to the one used for $1{\rm d}$ function approximation replaced by a two-dimensional input
$\bm{u}=(u_x,u_y)$, which is embedded through the collective-spin unitary

\begin{equation}
U_{\rm em}(\bm{u})=\exp\left[-i\left(u_x J_x+u_y J_y\right)\right].
\label{eq:emb_spiral}
\end{equation} 

From the main text it is clear that
the corresponding decision boundaries are not a robust representation of the underlying spiral manifold, indicating that the model cannot learn a robust structure for the underlying spiral manifold. Such limitation can be understood from the restricted Fourier structure generated by the collective embedding. Writing $u_x=r\cos\phi$ and $u_y=r\sin\phi$, one obtains $u_xJ_x+u_yJ_y =r(\cos\phi J_x+\sin\phi J_y)=r e^{-i\phi J_z}J_x e^{i\phi J_z}$ and therefore
\begin{equation}
U_{\rm em}(r,\phi)=
e^{-i\phi J_z}e^{-irJ_x}e^{i\phi J_z}.
\end{equation}
Hence, after the fixed encoder and decoder, each measurement probability is a matrix element of a finite-dimensional collective-spin representation,
$p_m(r,\phi)=\left|\langle m|U_{\rm de}e^{-i\phi J_z}e^{-irJ_x}e^{i\phi J_z}U_{\rm en}|\psi_0\rangle\right|^2.$
Since $J_z$ is diagonal in the Dicke basis, the angular dependence can only appear through a finite Fourier expansion,
$p_m(r,\phi)=
\sum_{q,q'}
A^{(m)}_{qq'}(r)e^{i(q-q')\phi}.
$
Consequently, the linear readout constructs decision functions of the form
\begin{equation}
F(r,\phi)=\sum_m w_m p_m(r,\phi)=\sum_{\ell=-2J}^{2J}F_\ell(r)e^{i\ell\phi}.
\end{equation}
The accessible classifier is therefore band-limited in the angular variable, with radial envelopes $F_\ell(r)$ constrained by the single collective rotation $e^{-irJ_x}$.

This structure is not naturally matched to a spiral decision boundary with a correlated radial-angular dependence of the form
$\phi\simeq ar+\phi_0$,
or equivalently Fourier components such as $\cos(\phi-ar)=\frac{1}{2}\left(e^{i\phi}e^{-iar}+e^{-i\phi}e^{iar}\right).$ Thus, a faithful representation of the spiral requires angular Fourier modes whose phases change progressively as a function of the radius. The collective embedding does not provide independently tunable radial-angular Fourier components of this type. It can therefore achieve moderate accuracy on the sampled data, especially for the one-turn case, but the resulting decision boundaries need not robustly reproduce the underlying spiral manifold. This limitation becomes more pronounced for the two-turn dataset, where the required radial-angular winding frequency is larger and the accessible collective-spin feature space is too rigid to represent the spiral geometry faithfully. 

On the other hand by replacing the permutation-symmetric collective embedding by the trainable local embedding of Eq.~\eqref{eq:emb_spiral2} allows different qubits to couple differently to the input coordinates and thus capture the correct boundary decisions.

To further probe the role of the input embedding, we also considered an intermediate architecture that lies between the fully collective embedding and the fully local trainable embedding of the main text. In this variant, the two input coordinates are encoded into two different halves of the spin register through collective $J_z$ rotations,

\begin{equation}
\label{eq:emb_spiral3}
U_{\rm em}(\mathbf{u})=\exp\left[-i\left(u_x J_z^{(A)} + u_y J_z^{(B)} \right) \right],
\end{equation}
where
\begin{equation}
J_z^{(A)}=\frac{1}{2} \sum_{i=1}^{L/2} \hat{\sigma}_z^{(i)}, \qquad J_z^{(B)}=\frac{1}{2}
\sum_{i=L/2+1}^{L} \hat{\sigma}_z^{(i)}.
\end{equation}

For odd $L$, the two blocks are chosen as two non-empty partitions of the spin register. This embedding treats the two input coordinates asymmetrically at the level of the spin register: $(u_x)$ is encoded in the collective magnetization of the first block, while $(u_y)$ is encoded in the collective magnetization of the second block. As a result, the embedding no longer preserves the full permutation symmetry of the register and therefore cannot, in general, be simulated within the fully symmetric $J=L/2$ subspace. The simulations for this embedding are consequently performed in the full $2^L$-dimensional Hilbert space, with computational-basis probabilities used as the measured features.

The resulting decision boundaries are presented in Fig.~\ref{fig:spiral_infinite_shots3}. Using similar training resources as in the previous local embedding we achieve an accuracy of $0.94$ and $0.71$ for the one-turn and two-turn dataset respectively and in the computational sensing approach with infinite shots. In the indirect approach we respectively achieve an accuracy of $0.91$ and $0.6$. We highlight that the reduced accuracy of the indirect approach on the two-turn dataset should not be interpreted as a failure of the circuit to encode the input coordinates. Indeed, the pairs $(u_x,u_y)$ can be estimated with very small error. Rather, the limitation arises in the subsequent post-processing step: a linear map from the reconstructed coordinates, or from the corresponding features, is not sufficiently expressive to implement the highly nonlinear decision rule required by the two-turn spiral.

\begin{figure}[t!]
\centering{}
\includegraphics[width=0.48\textwidth]{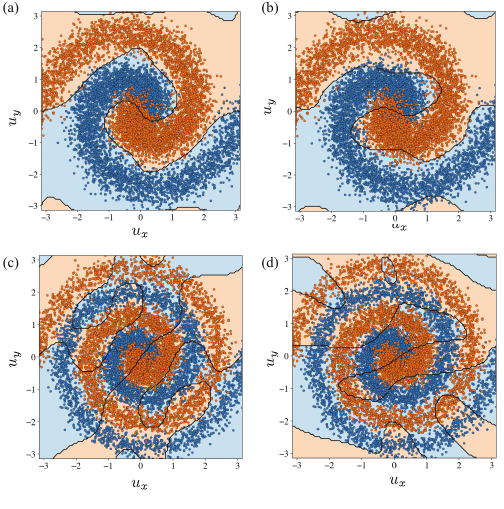} 
\caption{Binary classification of the noisy``two spirals'' dataset for the local embedding of Eq.~\ref{eq:emb_spiral3} with $L=6$, $n_{\rm en}=1$ and $n_{\rm de}=2$ with infinite shots. The background colors indicate the predicted class, revealing the learned decision boundaries. Panels (a,c) correspond to the direct (end-to-end, resource-aware) optimization approach, while panels (b,d) use the indirect approximation strategy. The top row (a,b) shows the one-turn spiral dataset, and the bottom row (c,d) the more challenging two-turn case. 
The direct approach achieves classification accuracy of $0.94$ and $0.71$ for the one- and two-turn datasets, compared to $0.91$ and $0.60$ obtained with the indirect method respectively. Note that the illustrated testing dataset is approximately ten times larger than the training one.}  
\label{fig:spiral_infinite_shots3} 
\end{figure}
\end{widetext}

\end{document}